\documentclass[%
 reprint,
superscriptaddress,
 amsmath,amssymb,
 aps,
 twocols,nofootinbib
]{revtex4-1}
\usepackage{verbatim}
\usepackage{graphicx}
\usepackage{dcolumn}
\usepackage{bm}
\usepackage{mathtools}  
\usepackage{booktabs}                              
           
\usepackage{float}                                    
      
\usepackage{graphicx}                                 
\usepackage{caption}                                 
\usepackage{subcaption}                               
\usepackage{hyperref}        
\usepackage{multirow}
\usepackage{hhline}                         
\usepackage[usenames, dvipsnames]{color}
\usepackage{color}
\usepackage{xr}

\newcolumntype{P}[1]{>{\centering\arraybackslash}p{#1}}

\begin{document}

\preprint{APS/123-QED}

\title{A mapping space Odyssey: characterising the statistical and metric properties\\ of reduced representations of macromolecules}

\author{Roberto Menichetti}
\affiliation{Physics Department, University of Trento, via Sommarive, 14 I-38123 Trento, Italy}
\affiliation{INFN-TIFPA, Trento Institute for Fundamental Physics and Applications, I-38123 Trento, Italy}
\author{Marco Giulini}
\affiliation{Physics Department, University of Trento, via Sommarive, 14 I-38123 Trento, Italy}
\affiliation{INFN-TIFPA, Trento Institute for Fundamental Physics and Applications, I-38123 Trento, Italy}
\author{Raffaello Potestio}%
 \email{raffaello.potestio@unitn.it}
\affiliation{Physics Department, University of Trento, via Sommarive, 14 I-38123 Trento, Italy}
\affiliation{INFN-TIFPA, Trento Institute for Fundamental Physics and Applications, I-38123 Trento, Italy}

\date{\today}

\begin{abstract}
Simplified representations of macromolecules help in rationalising and understanding the outcome of atomistic simulations, and serve to the construction of effective, coarse-grained models. The number and distribution of coarse-grained sites bears a strict relation with the amount of information conveyed by the representation and the accuracy of the associated effective model; in this work, we investigate this relationship from the very basics: specifically, we propose a rigorous notion of scalar product among mappings, which implies a distance and a metric space of simplified representations. Making use of a Wang-Landau enhanced sampling algorithm, we exhaustively explore the space of mappings, quantifying their qualitative features in terms of their squared norm and relating them with thermodynamical properties of the underlying macromolecule. A one-to-one correspondence with an interacting lattice gas on a finite volume leads to the emergence of discontinuous phase transitions in mapping space that mark the boundaries between qualitatively different representations of the same molecule.\end{abstract}

\pacs{Valid PACS appear here}

\maketitle

\section{Introduction}

The research area of computational molecular biophysics has experienced, in the past few decades, impressive advancements in two complementary and strictly intertwined fields: on the one hand, the steadily growing and increasingly cheaper computational power has enabled the simulation of ever larger systems with atomistic resolution \cite{SINGHAROY20191098, zimmerman2021sars}; on the other hand, there has been an explosion of diverse coarse-grained (CG) models \cite{Takada2012, noid_persp, kmiecik2016coarse}, i.e. simpler representations of molecules in terms of relatively few sites interacting through effective potentials: these have filled several gaps between the length- and time-scales of interest and the current capability of all-atom methods to cover them. The scientific efforts making use of one or both these techniques have cracked several important problems open, ranging from protein folding to cell growth.

The development of a successful CG model is strongly dependent on the choice of the reduced representation, or CG mapping, and on the correct parametrization of the effective interactions \cite{noid_persp,giulini2021system}. The latter challenge has received an enormous amount of attention, leading to extremely accurate and sophisticated procedures to determine approximate CG potentials such as OPEP \cite{maupetit2007coarse}, PRIME \cite{voegler2001alpha} and UNRES \cite{liwo2014unified}. The former task has been the object of a smaller number of works, however its centrality in and beyond the process of coarse graining has recently started to emerge \cite{foley2020exploring, giulini2021system}; indeed, a deep relationship exists between the degrees of freedom one selects to {\it construct} a CG model of the system, and those one employs to {\it analyse} its behaviour from a more detailed representation.

On the one hand, high-resolution, fully atomistic models are necessarily required to let the properties and behaviour of complex biomolecular systems emerge; on the other hand, the interpretation and understanding of this behaviour requires a reduction of the mountain of {\it data} and its synthesis in a smaller amount of {\it information}. In a nutshell, while the generative process has to be high-resolution to be useful, its outcome has to be low-resolution to be intelligible. An intuitive example of this concept is given by the representation of a protein structure in terms of its $C_{\alpha}$'s, i.e. the alpha carbons of the backbone: this mapping is not only extensively employed in the development of CG models \cite{clementi2000topological, atilgan2001anisotropy}(e.g., the whole amino acid is represented as a single bead whose position coincides with that of the C$_\alpha$), but it is also extremely common in the analysis of structures sampled in fully atomistic simulations \cite{lindorff2011fast, grottesi2020computational}.

A few different strategies have been developed that aim at identifying the optimal CG mapping to describe a molecule, which differ most notably in the observable used to drive the optimization. There exists a first class of algorithms that rely on a completely static, graph-based description of the system \cite{delvenne2010stability, depabloJCTC2019}, while a second group of approaches makes use of the dynamics of the system, obtained through models with more \cite{Wang_2019, giulini2020information} or less \cite{foley2015impact, potestio_jctc} detailed force fields. For instance, a recent protocol proposed by us \cite{giulini2020information} revolves around the analysis of an all-atom molecular dynamics (MD) \cite{md_general_method, md_sim_biomol} simulation trajectory of a protein in terms of a subset of the molecule's atoms; a physics-driven choice of the latter allows one to identify the one or few mappings that return the most parsimonious yet informative simplified description of the system.

It is in this context that several questions arise, which, before tackling the issues related to the {\it properties} that a mapping can let emerge, pertain the mapping itself, specifically: how many mappings are there? How many of them are {\it interesting} (whatever this adjective implies)? What is the number of available mappings for a given number of selected atoms? How can we quantify the difference, or {\it distance}, between two mappings?

The answers to these questions have twofold importance. On the one hand, they provides us with the mathematical and computational tools required to make the most of those analysis methods that rely on the concept of mapping; on the other hand, they bring to the surface a large amount of interesting physics that emerges from the analysis of the reduced representations of a single structure, which can be then employed to rationalise the properties of molecules when observed in simpler terms.

This work takes the moves from the introduction of a mathematically rigorous notion of distance between mappings of a given molecule. This apparently simple object constitutes the bedrock of subsequent analyses, in that it enables the exploration of the metric space induced by this distance and the associated scalar product. Through the application of an enhanced sampling algorithm, namely the Wang-Landau method \cite{wang2001determining, wang2001efficient}, we characterise the mapping space, and associate its properties to structural features of the underlying molecule. Finally, the isomorphism between the problem of exploring the possible mappings of a molecule and that of a lattice gas in a finite volume shows the emergence of first-order phase transitions in the latter, distinguishing mappings with qualitatively different properties.

The paper is organised as follows: in Sect. \ref{sec:theory} we develop the scalar product between decimation mappings, and derive from it a notion of distance in mapping space; in Sect. \ref{sec:explore} we study the mappings in terms of the distribution of values of the squared norm of mappings having a given number of retained sites $N$, first through random sampling, then making use of the Wang-Landau enhanced sampling method; in Sect. \ref{sec:phasetrans} we exploit a duality between the problem of mappings of a macromolecule and that of an interacting lattice gas in a finite volume to investigate the properties of mappings; in Sect. \ref{sec:topology} we investigate the topology of the mapping space making use of the distance between reduced representations, which enables a low-dimensional representation that highlights the general features of this space; in Sect. \ref{sec:topology} we sum up the results of this work and discuss its future perspectives.

\section{Theory}
\label{sec:theory}

The construction of a CG model for a macromolecular system starts with the selection  of a \emph{mapping} $M$, that is, the projection operator connecting a  microscopic, detailed configuration ${\bf r}_i,\;i=1,...,n$ to a low-resolution one ${\bf R}_I,\; I=1,...,N<n$,
\begin{eqnarray}
\label{eq:mapping_symb}
&M& = \{ {\bf M}_I({\bf r}),\;\; I=1,...,N \}, \nonumber\\
&&{\bf M}_I({\bf r})={\bf R}_I=\sum_{i=1}^n c_{Ii} {\bf r}_i,
\end{eqnarray}
where $n$ and $N$ are the number of atoms in the system and the number of effective interaction sites employed in its CG simplified picture, respectively.
In Eq.~\ref{eq:mapping_symb}, the weights $c_{Ii}$ are positive, spatially homogeneous---i.e. independent of the configuration ${\bf r}$---and subject to the normalization condition $\sum_{i=1}^n c_{Ii}=1$ to preserve translational invariance \cite{noid_persp}. While a particular choice of these coefficients corresponds to a specific CG representation of the system, by varying them, along with changing the degree of CG'ing $N$, one spans the \emph{mapping space} $\mathcal{M}$, whose elements are all the possible low-resolution descriptions that can be assigned to a macromolecule.

In the perspective of quantitatively characterizing the properties of such space, the cardinality of $\mathcal{M}$ in the continuous definition presented in Eq.~\ref{eq:mapping_symb} makes its thorough exploration, although appealing, hard to handle in practice. In this work, we thus restrict our analysis to the \emph{discrete} subspace of CG representations that can be obtained for a system through a decimation of its microscopic degrees of freedom: a subset of $N$ constituent atoms is retained while the remaining ones are neglected. Despite the simplifications introduced by this procedure, the number of CG representations $\Omega_N$ that can be constructed for a macromolecule selecting $N$ atoms out of $n$ is
\begin{equation}
\label{eq:firstomegaN}
\Omega_N=\frac{n!}{N!(n-N)!},
\end{equation} 
so that the \emph{total} number of possible decimation mappings $\Omega$ reads
\begin{equation}
\Omega=\sum_{N=1}^n \Omega_N=\sum_{N=1}^n \frac{n!}{(n-N)!N!}=2^n -1,
\end{equation}
which becomes prohibitively large as the size of the system increases. Consequently, in the following we only consider the heavy atoms of the molecule as candidate CG sites, indicating with $\mathcal{M}$ the subspace of CG mappings obtained according to these prescriptions.

The investigation of the topological structure of $\mathcal{M}$ calls for the introduction of a distance $\mathcal{D}(M,M'),\ M,M' \in \mathcal{M}$, able to quantify the ``separation'' between pairs of points $M$ and $M'$ belonging to the space of decimation mappings, that is, pairs of CG representations employed to represent the system that differ in the choice of the retained atoms. Such distance must be equipped with all the associated metric properties, namely identity, symmetry, and triangle inequality.

To construct $\mathcal{D}(M,M')$, we consider a \emph{static} configuration of the system with (heavy) atoms located in positions $\mathbf{r}_i,~i=1,...,n$ and a set of selection operators $\chi_{M,i},~i=1,..,n$ defining mapping $M$,
\begin{eqnarray}
\label{eq:mapping}
\chi_{M,i} &&=
\left\{
	\begin{array}{ll}
		1  & \mbox{if atom $i$ is retained,}\\
		0 & \mbox{if atom $i$ is not retained,}
	\end{array}
\right.\\
\label{eq:mapping_degree}
&&\;\;\;\;\;\;\;\;\sum_{i = 1}^n \chi_{M,i} = N(M),
\end{eqnarray}
where $N(M)$ is the number of retained atoms in the mapping. Taking inspiration from the Smooth Overlap of Atomic Positions method (SOAP) developed by Cs\'any \emph{et al.} \cite{bartok2013representing,de2016comparing}, we associate to each $M\in\mathcal{M}$ an element $\phi_M({\bf r})$ of the Hilbert space of square-integrable real functions $L_2(\mathbb{R}_3)$ as
\begin{equation}
\label{eq:sitegaus}
\phi_M(\mathbf{r})=\sum_{i=1}^n\phi_{M,i}(\mathbf{r})=\sum_{i=1}^n Ce^{-({\mathbf r}-{\mathbf r}_i)^2/2\sigma^2}\chi_{M,i},
\end{equation}
obtained by centering a three-dimensional Gaussian---whose normalization factor $C$ will be fixed in the following---on the position of each atom of the macromolecule retained in the mapping.\footnote{In contrast to the original definition of the SOAP measure---which enables to quantify the similarity between two molecular structures \cite{bartok2013representing,de2016comparing}---we here aim at determining the overlap between different CG representations of a \emph{single} compound. As such, with respect to SOAP: \emph{(i)} in Eq.~\ref{eq:sitegaus} we do not employ local densities representing the chemical environment of a \emph{specific} atom (which would afterwards require, e.g., to average over all pairs of atoms for the calculation of the total similarity kernel \cite{de2016comparing}), but rather global ones associated to the molecule as a whole; and \emph{(ii)} in Eq.~\ref{eq:dotprod} we do not introduce an additional integral over rotations of one of the two structures. Indeed, there is no ambiguity in defining the alignment of different CG representations, as this is dictated by the original, full-atom reference.}

The inner product $\langle\phi_M,\phi_{M'}\rangle$ of $L_2(\mathbb{R}_3)$ between two mappings $M$ and $M'$,
\begin{equation}
\label{eq:dotprod}
\langle \phi_M,\phi_{M'}\rangle=\int d\mathbf{r}\ \phi_M(\mathbf{r})\phi_{M'}(\mathbf{r}),
\end{equation}
induces a norm $||\phi_M||$ for mapping $M$, with
\begin{equation}
\label{eq:norm}
\mathcal{E}(M)=||\phi_M||^2=\langle \phi_M,\phi_M\rangle,
\end{equation}
starting from which the distance $\mathcal{D}(M,M')$ can be defined as
\begin{eqnarray}
\label{eq:distdot}
&&\mathcal{D}(M,M')=||\phi_M-\phi_{M'}||\nonumber \\
&&=\langle \phi_M-\phi_{M'},\phi_M-\phi_{M'}\rangle^{\frac{1}{2}},
\end{eqnarray}
$\mathcal{D}(M,M')$ satisfying  all the aforementioned metric properties.

By inserting Eq.~\ref{eq:sitegaus} in Eq.~\ref{eq:dotprod}, the inner product $\langle \phi_M, \phi_{M'}\rangle$ between mappings generated by two distinct selection operators $\chi_{M}$ and $\chi_{M'}$ becomes
\begin{equation}
\label{eq:dotmapp}
\langle \phi_M, \phi_{M'}\rangle=\sum_{i,j=1}^n J_{ij}\chi_{M,i}\chi_{M',j},
\end{equation}
while the associated distance $\mathcal{D}(M,M')$  in Eq.~\ref{eq:distdot} reads
\begin{eqnarray}
\label{eq:dist_dotmapp}
&&\mathcal{D}(M,M')= \left(\mathcal{E}(M)+\mathcal{E}(M')-2\langle \phi_M,\phi_{M'}\rangle\right)^{\frac{1}{2}}\nonumber \\
&&=\left( \sum_{i,j=1}^n J_{ij}\chi_{M,i}\chi_{M,j} \;+\; \sum_{i,j=1}^n J_{ij}\chi_{M',i}\chi_{M',j}\;+\right. \nonumber\\ 
&&\;\;\;\left. -2\sum_{i,j=1}^n J_{ij}\chi_{M,i}\chi_{M',j}\right)^{\frac{1}{2}}.
\end{eqnarray}
In Eq.~\ref{eq:dotmapp} and \ref{eq:dist_dotmapp}, the coupling constant $J_{ij}=J_{ij}(\mathbf{r}_i,\mathbf{r}_j)$ between two atoms $i$ and $j$ is given by
\begin{equation}
\label{eq:jijdef}
J_{ij}(\mathbf{r}_i,\mathbf{r}_j)=C^2\int d\mathbf{r}\ e^{-[(\mathbf{r}-\mathbf{r}_i)^2+(\mathbf{r}-\mathbf{r}_j)^2]/2\sigma^2},
\end{equation}
with
\begin{equation}
J_{ij}(\mathbf{r}_i,\mathbf{r}_j)=J_{ij}(|\mathbf{r}_i-\mathbf{r}_j|)=J_{ij}(r_{ij}).
\end{equation}
due to translational and rotational invariance. By introducing polar coordinates in Eq.~\ref{eq:jijdef}, one has
\begin{eqnarray}
\label{eq:couplconst}
&&J_{ij}(r_{ij})=2\pi C^2\int dr d\theta\ r^2 \sin\theta e^{-\frac{1}{2\sigma^2}(2r^2+r_{ij}^2-2rr_{ij}\cos\theta)}\nonumber \\ 
&&=\frac{4\pi\sigma^2}{r_{ij}}C^2 e^{-r^2_{ij}/2\sigma^2}\int dr\ re^{-r^2/\sigma^2}\sinh\left(\frac{rr_{ij}}{\sigma^2}\right),
\end{eqnarray}
and a chain of Gaussian integrals provides
\begin{equation}
\label{eq:jijform}
J_{ij}(r_{ij})=\pi^{3/2}C^2\sigma^3 e^{-r^2_{ij}/4\sigma^2}=e^{-r^2_{ij}/4\sigma^2},
\end{equation}
where the last equality has been obtained by setting, without loss of generality,
\begin{equation}
 C^2=\frac{1}{\pi^{3/2}\sigma^3}.
\end{equation}
Finally, by combining Eq.~\ref{eq:dotmapp} and \ref{eq:jijform} the inner product $\langle \phi_M,\phi_{M'}\rangle$ reads
\begin{equation}
\label{eq:scal_prod}
\langle \phi_M,\phi_{M'}\rangle=\sum_{i,j=1}^n e^{-r^2_{ij}/4\sigma^2}\chi_{M,i}\chi_{M',j},
\end{equation}
i.e. a sum of Gaussian factors over the positions of all pairs of atoms retained in the two mappings. Notably, the factorization with respect to the operators $\chi_{M}$ and $\chi_{M'}$ in Eq.~\ref{eq:dotmapp} and \ref{eq:scal_prod} enables the inner product (and therefore the distance $\mathcal{D}$ and the squared norm $\mathcal{E}$) to be determined starting from a  matrix $J_{ij}$ that can be calculated \emph{a priori} over the static structure of the molecule.

\begin{figure*}[pt]
		\includegraphics[width=\textwidth]{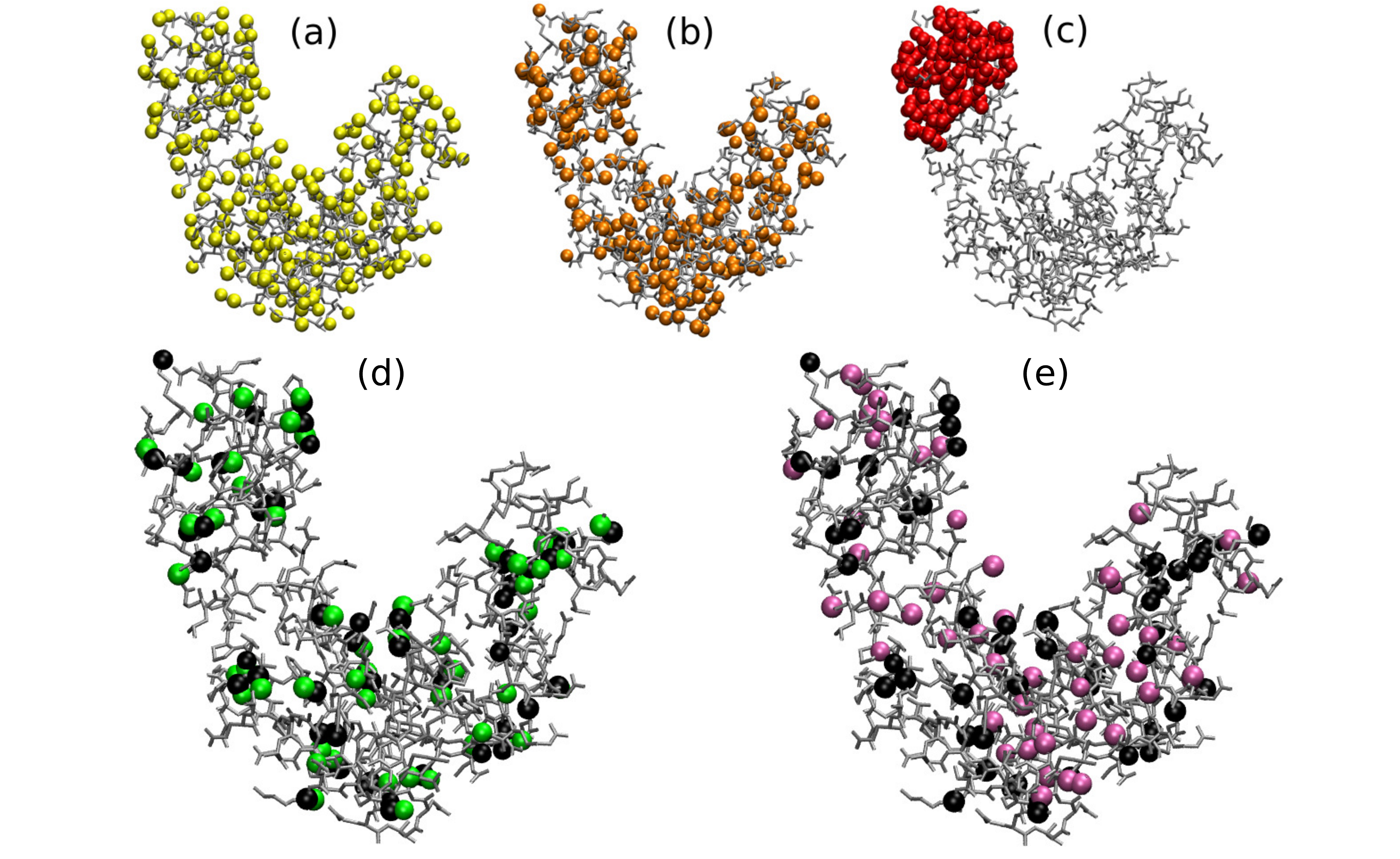}
		\caption{\emph{Top row}: Example of possible CG representations for adenylate kinase with $N=214$ sites (represented as beads) characterised by a low (a), intermediate (b) and high (c) mapping squared norm $\mathcal{E}$. By increasing $\mathcal{E}$ we move from maximally homogeneous to extremely globular CG representations. \emph{Bottom row}: Examples of CG mappings with $N=53$ sites that are approximately parallel (d) and orthogonal (e) to a given one. The atoms composing the reference CG representation are represented as black beads. Parallel (resp. orthogonal) mappings tend to displace CG sites on similar (resp. complementary) regions of the system. \label{fig:mappings}}
\end{figure*}

One might ask what kind of information the previously defined quantities provide about the possible CG representations of a system. To answer this question, we first focus on the squared norm of a mapping $\mathcal{E}(M)$, see Eq.~\ref{eq:norm} and \ref{eq:scal_prod},
\begin{equation}
\label{eq:norm_gauss}
\mathcal{E}(M)=\langle \phi_M,\phi_{M}\rangle=\sum_{i,j=1}^n e^{-r^2_{ij}/4\sigma^2}\chi_{M,i}\chi_{M,j}.
\end{equation}
For a given retained atom $i$, the sum over $j$ in Eq.~\ref{eq:norm_gauss},
\begin{equation}
Z_i(M)=\sum_{j=1}^n e^{-r^2_{ij}/4\sigma^2}\chi_{M,j},
\end{equation}
approximately represents its CG coordination number, that is, the number of retained atoms in the mapping that are located within a sphere of radius $\sqrt{2}\sigma$ from $i$. By fixing the degree of coarse-graining $N$, $\mathcal{E}(M)$ scales as 
\begin{eqnarray}
\label{eq:scal_norm}
\mathcal{E}(&&M) = N\bar{Z}(M),\\
\bar{Z}(M)&&=\frac{1}{N}\sum_{i=1}^n Z_i(M)\chi_{M,i}
\end{eqnarray}
showing that the dependence of the norm on the specific selection of atoms is dictated by $\bar{Z}(M)$, the \emph{average} CG coordination number. Let us now consider two limiting cases: \emph{(i)} extremely sparse and homogeneous CG representations, in which each retained atom does not have any retained neighbour within a radius of order $\sigma$---this condition can only be fulfilled provided that $N$ is not too large, \emph{vide infra}, or $\sigma$ is much smaller than the typical interatomic distance. In this case, one has $\bar{Z}(M)\approx 1$ and consequently $\mathcal{E}(M)\approx N$; \emph{(ii)} globular mappings characterized by densely populated (i.e. almost atomistic) regions of retained sites surrounded by ``empty'' ones. In this case, the average coordination number $\bar{Z}(M)$ will roughly resemble its atomistic counterpart, the latter being defined as
\begin{equation}
\label{eq:at_coord}
\bar{z}=\frac{1}{n}\sum_{i,j=1}^n e^{-r^2_{ij}/4\sigma^2},
\end{equation} 
and thus $\mathcal{E}(M)\approx N\bar{z}$. It follows that the squared norm $\mathcal{E}(M)$ captures the average homogeneity of a CG representation, that is, whether the associated retained atoms are uniformly distributed across the macromolecule or are mainly localized in well-defined regions of it. In Fig.~\ref{fig:mappings} we report examples of CG mappings extracted for these two extreme categories in the case of adenylate kinase (see Sec. \ref{sec:explore} for further details on this protein) together with a CG representation in which the retained atoms are randomly selected.

An analogous discussion can be performed for the inner product $\langle \phi_M,\phi_{M'}\rangle$ in Eq.~\ref{eq:scal_prod}, calculated between two mappings $M$ and $M'$ respectively retaining $N$ and $N'$ atoms of the system. For a given atom $i$ in mapping $M$,
\begin{equation}
\label{eq:T_single}
T_i(M')=\sum_{j=1}^n e^{-r^2_{ij}/4\sigma^2}\chi_{M',j}
\end{equation}
approximately counts the number of neighbours $j$ in mapping $M'$ located within a sphere of radius $\sqrt{2}\sigma$ from $i$. The inner product scales as
\begin{eqnarray}
\label{eq:dot_T}
\langle \phi_M,\phi_{M'}\rangle &&= N\bar{T}(M,M'),\\
\bar{T}(M,M')&&=\frac{1}{N}\sum_{i=1}^n T_i(M')\chi_{M,i},
\label{eq:T_def}
\end{eqnarray}
where $\bar{T}(M,M')$ is again the \emph{average} number of neighbours an atom in mapping $M$ has that belong to mapping $M'$. Eqs.~\ref{eq:T_single}, \ref{eq:dot_T} and \ref{eq:T_def} provide a very intuitive explanation of the orthogonality of mappings, $\langle \phi_M,\phi_{M'}\rangle\approx 0$: it is sufficient that each atom in mapping $M$ does not have any neighbour in $M'$ (and obviously vice-versa). As such, orthogonal mappings cover complementary regions of the system. 

In general, the existence of an inner product enables the definition of an angle $\theta_{M,M'}$ between mappings, whose cosine reads
\begin{equation}
\label{eq:cosine}
\cos\theta_{M,M'}=\frac{\langle \phi_M,\phi_{M'}\rangle}{\left(\mathcal{E}(M)\mathcal{E}(M')\right)^{\frac{1}{2}}}.
\end{equation}
While the orthogonality of mappings ($\cos\theta_{M,M'}\approx 0$) has a relatively straightforward interpretation in terms of their spatial complementarity, the condition of parallelism, $\cos\theta_{M,M'}\approx 1$, is a bit less intuitive. If the mappings $M$ and $M'$ have the same number of atoms $N$, by inserting Eq.~\ref{eq:scal_norm} and \ref{eq:dot_T} in Eq.~\ref{eq:cosine} one obtains
\begin{equation}
\cos\theta_{M,M'}=\frac{\bar{T}(M,M')}{\left(\bar{Z}(M)\bar{Z}(M')\right)^{\frac{1}{2}}}.
\end{equation}
If furthermore the two mappings show also roughly the same ``globularity'', $\bar{Z}(M)\approx\bar{Z}(M')$, their parallelism requires
\begin{equation}
\bar{T}(M,M')\approx\bar{Z}(M),
\end{equation}
that is, the average number of neighbors one atom of $M$ has from mapping $M'$ has to be equal to the average number of neighbors the has from itself. This means that the two mappings must place retained atoms across the macromolecule in a similar fashion. Examples of parallel and orthogonal CG representations for adenylate kinase are presented in Fig~\ref{fig:mappings}.

It follows that while $\mathcal{E}(M)$ quantifies the average sparseness of a CG representation, $\langle \phi_M,\phi_{M'}\rangle$---or equivalently $\cos\theta_{M,M'}$---characterizes the average degree of spatial similarity between two different decimations of the microscopic degrees of freedom of the system. The distance $\mathcal{D}(M,M')$ in Eq.~\ref{eq:dist_dotmapp} combines these two notions to extract how ``far'' a pair of CG representations is in the space of possible mappings $\mathcal{M}$.  

Based on these observations, we implemented a slight modification to the inner product $\langle \phi_M,\phi_{M'}\rangle$---and hence to the squared norm $\mathcal{E}(M)$ and distance $\mathcal{D}(M,M')$---with respect to the definition originally presented in Eq.~\ref{eq:scal_prod}, which however does not change its overall properties or interpretation. We have previously discussed how in the limiting cases of extremely sparse and globular mappings one respectively obtains $\mathcal{E}(M)\approx N$ and $\mathcal{E}(M)\approx  N\bar{z}$, where $\bar{z}$ is the atomistic coordination number in Eq.~\ref{eq:at_coord}. As the number of CG sites $N$ increases, however, it will be extremely hard for a retained site not to have any retained neighbor within a sphere of radius of order $\sigma$, so that the exact scaling of $\mathcal{E}(M)$ on the degree of CG'ing $N$ in the case of sparse mappings will be hardly observed. We thus divide the inner product in Eq.~\ref{eq:scal_prod} by the average atomistic coordination number, and define
\begin{equation}
\label{eq:scal_rescaled}
\langle \phi_M,\phi_{M'}\rangle_{\bar{z}}~=\frac{1}{\bar{z}}~\langle \phi_M,\phi_{M'}\rangle.
\end{equation}
Consequently, one has
\begin{eqnarray}
\label{eq:rescaled_norm}
\mathcal{E}_{\bar{z}}(M)~&&=\frac{1}{\bar{z}}~\mathcal{E}(M),\\
\mathcal{D}_{\bar{z}}(M,M')~&&=\frac{1}{\sqrt{\bar{z}}}~\mathcal{D}(M,M'),
\end{eqnarray}
while the cosine between two mappings $\cos\theta_{M,M'}$ is not affected by the rescaling.
With this choice, globular mappings are now associated to $\mathcal{E}(M)_{\bar{z}}\approx N$, which can always be observed also in the case of low degrees of CG'ing, that is, high $N$. Note that the definition of $\langle \phi_M,\phi_{M'}\rangle_{\bar{z}}$ in Eq.~\ref{eq:scal_rescaled} corresponds to a rescaling of the coupling constant $J_{ij}$ in Eq.~\ref{eq:jijform} to
\begin{equation}
J_{ij}=\frac{1}{\bar{z}}~e^{-r^2_{ij}/4\sigma^2}.
\end{equation}
For notational convenience, in the following we will omit the subscript $\bar{z}$ and refer to $\mathcal{E}(M)_{\bar{z}}$, $\langle \phi_M,\phi_{M'}\rangle_{\bar{z}}$ and $\mathcal{D}_{\bar{z}}(M,M')$ as $\mathcal{E}(M)$, $\langle \phi_M,\phi_{M'}\rangle$ and $\mathcal{D}(M,M')$, respectively.

\section{Exploration of the mapping space}
\label{sec:explore}

Starting from the definitions introduced in Sec.~\ref{sec:theory}, we now proceed to perform a quantitative analysis of the high-dimensional space $\mathcal{M}$ of CG representations that can be constructed for a macromolecule through a decimation of its atomistic degrees of freedom. As a testbed system we consider \emph{adenylate kinase} (AKE), a $214$ residue-long phosphotransferase enzyme catalysing the interconversion between adenine nucleotides, namely adenine diphosphate (ADP), adenine monophosphate (AMP), and the adenine triphosphate complex (ATP)~\cite{4AKE}. The structure of adenylate kinase can be divided in three main building blocks \cite{Pontiggia2008,Potestio2009}, with the mobile LID and NMP domains exhibiting a conformational rearrangement around a hinge, the stable CORE domain, which results in an overall \emph{open~$\leftrightarrow$~closed} transition of the enzyme~\cite{shapiro_2009,formoso_2015}. Our calculations require in input only a static configuration $\mathbf{r}_i,~i=1,...,n$ of the system to determine the set of Gaussian couplings $J_{ij}$ in Eq.~\ref{eq:jijform}. We here rely on the \emph{open} crystal conformation of adenylate kinase (PDB code 4AKE), excluding from the analysis all hydrogens composing the biomolecule, resulting in a total of 1656 heavy atoms.

The investigation of the topological structure of the decimation mapping space of AKE calls for an extensive characterisation of the relational properties among its points, achievable by analysing the behaviour of the distance $\mathcal{D}(M,M')$ over an ensemble of prototypical CG representations extracted from $\mathcal{M}$. The discussion carried out in Sec.~\ref{sec:theory}, however, highlighted that $\mathcal{D}(M,M')$ strictly depends on two factors: the globularity of each mapping---encoded in the squared norm $\mathcal{E}(M)$---and their mutual spatial complementarity---that is, the inner product $\langle \phi_M,\phi_{M'}\rangle$ or equivalently the cosine $\cos\theta_{M,M'}$. It is then useful to first focus on these one- and two-``body'' ingredients before combining them into the distance $\mathcal{D}(M,M')$. As such, in Sec.~\ref{sec:norm} and \ref{sec:cosine} we will respectively discuss the behaviour of $\mathcal{E}(M)$ and $\cos\theta_{M,M'}$ across the mapping space of AKE; the analysis of the distance $\mathcal{D}$, and hence of the topology of $\mathcal{M}$, will be presented in Sec.~\ref{sec:topology}.

\subsection{Norm distributions}
\label{sec:norm}

Let us first consider the squared norm $\mathcal{E}(M)$ of a CG representation $M$ defined in Eq.~\ref{eq:rescaled_norm}. As previously discussed, this quantity provides information about the spatial homogeneity of a mapping with a given degree of CG'in $N$; that is to say, it recapitulates how the retained atoms are distributed across the molecular structure, from uniformly scattered ($\mathcal{E}(M)\approx N/\bar{z}$) to mainly concentrated in well-defined, almost atomistic domains emerging out of a severely CG'ed background ($\mathcal{E}(M)\approx N$). 

It is important to stress that mappings belonging to the two aforementioned extreme cases are routinely employed by the CG'ing community in the description of a biomolecular system. In proteins, examples from the homogeneous class include physically-intuitive, residue-based CG representations of the molecule in terms of its $\alpha$ carbons or backbone atoms~\cite{kmiecik2016coarse,giulini2021system};  homogeneity, on the other hand, is often abruptly broken in chemically-informed, \emph{multiscale} mappings, in which a higher level of detail, up to the atomistic one, is sharply localized on the biologically/chemically relevant regions of the system---e.g. the active sites of the protein---while the reminder is treated at extremely low resolution~\cite{giulini2021system}. Furthermore, moving away from these limiting cases, an increasing attention is being posed in employing CG descriptions in which the level of detail is, although inhomogeneously, quasi-continuosly modulated throughout the molecular structure~\cite{giulini2021system}.

Be they fully homogeneous, markedly inhomogeneous, or smoothly interpolating between these two classes, the CG representations that are usually adopted in the literature to simplify a biomolecule are often selected \emph{a priori} by relying on general and intuitive criteria. Critically, such representations only constitute elements, isolated instances extracted from the high-dimensional mapping space $\mathcal{M}$ of the system.
One natural question follows: how representative are these ``common'' mappings of the diversity of the space $\mathcal{M}$? In other words, how spatially homogeneous are the possible CG descriptions that can be designed for a macromolecule when no prior knowledge about its chemical structure or biological function is exploited to guide the mapping construction?

\begingroup
\setlength{\tabcolsep}{8pt}
\renewcommand{\arraystretch}{1.1}
\begin{table}
\begin{tabular}{c c c | c c |}
 & \multicolumn{2}{c|}{$\langle\mathcal{E}\rangle_N$} & \multicolumn{2}{c}{$\sigma_{\mathcal{E},N}$}\\
 \cline{2-5}
\multicolumn{1}{c|}{$N$} & RS & WL-SP & RS & WL-SP \\ 
\hhline{|=|==|==|}
\multicolumn{1}{|c|}{$53$} & 5.41 & ---&  0.31 & --- \\
\multicolumn{1}{|c|}{$107$} & 14.15 & ---& 0.63 & --- \\
\multicolumn{1}{|c|}{$214$} & 41.14 & 40.82 & 1.32 & 1.32 \\
\multicolumn{1}{|c|}{$321$} & 80.95 & ---& 2.03 & --- \\
\multicolumn{1}{|c|}{$428$} & 133.58 & 133.17 & 2.74 & 2.74 \\
\multicolumn{1}{|c|}{$535$} & 199.04 & ---& 3.45 & --- \\
\multicolumn{1}{|c|}{$642$} & 277.33 & 276.93 & 4.12 & 4.11 \\
\multicolumn{1}{|c|}{$749$} & 368.44 & ---& 4.74 & --- \\
\multicolumn{1}{|c|}{$856$} & 472.39 & 471.95 & 5.29 & 5.29 \\
\multicolumn{1}{|c|}{$963$} & 589.16 & ---& 5.74 & --- \\
\multicolumn{1}{|c|}{$1070$} & 718.76 & 718.29 & 6.06 & 6.07 \\
\multicolumn{1}{|c|}{$1177$} & 861.18 & ---& 6.22 & --- \\
\multicolumn{1}{|c|}{$1284$} & 1016.43 & 1016.14 & 6.16 & 6.17 \\
\multicolumn{1}{|c|}{$1391$} & 1184.51 & ---& 5.79 & --- \\
\multicolumn{1}{|c|}{$1498$} & 1365.42 & 1365.05 & 4.94 & 4.94 \\
\multicolumn{1}{|c|}{$1605$} & 1559.15 & ---& 3.09 & --- \\
\hline
\end{tabular}
\caption{\label{tab:RSvsSP} Average mapping squared norm $\langle\mathcal{E}\rangle_N$ and associated standard deviation $\sigma_{\mathcal{E},N}$ at different degrees of coarse-graining $N$, calculated over the mapping space $\mathcal{M}$ of AKE. We present random sampling results (RS), as well as those obtained from a saddle-point approximation to the density of states $\Omega_N(\mathcal{E})$ determined through the Wang-Landau method (WL-SP), see text.}
\end{table}
\endgroup

To answer this question, we start by introducing the number of mappings that attain a particular value $\mathcal{E}$ of the squared norm for a given number of CG sites $N$, which is given by:
\begin{eqnarray}
\label{eq:omegaE}
&&\Omega_N(\mathcal{E}) = \sum_{M\in \mathcal{M}}\delta(N(M),N)\delta({\mathcal{E}(M),\mathcal{E}})
\end{eqnarray}
with
\begin{eqnarray}
\label{eq:omegaE2}
&&\sum_{M\in \mathcal{M}}\mathcal{O}(M)=\sum_{\chi_1=0,1}...\sum_{\chi_n=0,1}\mathcal{O}(\lbrace\chi_i\rbrace),
\end{eqnarray}
where $\mathcal{O}$ is a generic observable that depends on the mapping through the operators $\chi_i$. Normalizing Eq.~\ref{eq:omegaE} by the total number of mappings with $N$ sites, $\Omega_N$, we define the {\it probability} of having a mapping with given $\mathcal{E}$ and $N$, that is:
\begin{equation}
\label{eq:pgiven}
P_N(\mathcal{E}) = \frac{\Omega_N(\mathcal{E})}{\Omega_N},
\end{equation}
which satisfies the normalization condition
\begin{equation}
\sum_{\mathcal{E}}P_N(\mathcal{E})=1
\end{equation}
regardless of the number of retained sites. $P_N(\mathcal{E})$ can be rewritten as
\begin{equation}
P_N(\mathcal{E})=\left(\frac{n!}{(n-N)!N!}\right)^{-1}\sideset{}{'}\sum_{M\in \mathcal{M}}\delta(\mathcal{E}(M),\mathcal{E}),
\end{equation}
where the primed sum runs over all mappings with fixed resolution $N$, i.e. over all values of the set of operators $\chi_i=0,1,~i=1,..,n$ satisfying
\begin{equation}
\label{eq:normalization}
\sum_{i = 1}^n \chi_i = N.
\end{equation}

By providing direct insight on the degree of spatial uniformity characterising the ensemble of all possible CG descriptions of a macromolecular system, $P_N(\mathcal{E})$ represents a first important ingredient in the investigation of the structure of the mapping space $\mathcal{M}$. We thus aimed at investigating the behaviour of the conditional probability $P_N(\mathcal{E})$ across the decimation mapping space $\mathcal{M}$ of AKE for a set of $16$ values of $N$ ranging from $N=53$ to $1605$, see Table~\ref{tab:RSvsSP}. However, even restricted to these cases, an exhaustive enumeration of all possible CG representations of the system is unfeasible in practice: for example, in the case of AKE ($n=1656$), roughly $10^{276}$ possible CG representations can be constructed that describe the enzyme in terms of a subset of $N=214$ heavy atoms (one for each residue). This number grows to $10^{496}$ for $N=856$ (four heavy atoms per residue on average), that is, close to the maximum of the binomial coefficient, obtained for $N=n/2$, see Eq.~\ref{eq:firstomegaN}.

\begin{figure}
		\includegraphics[width=\columnwidth]{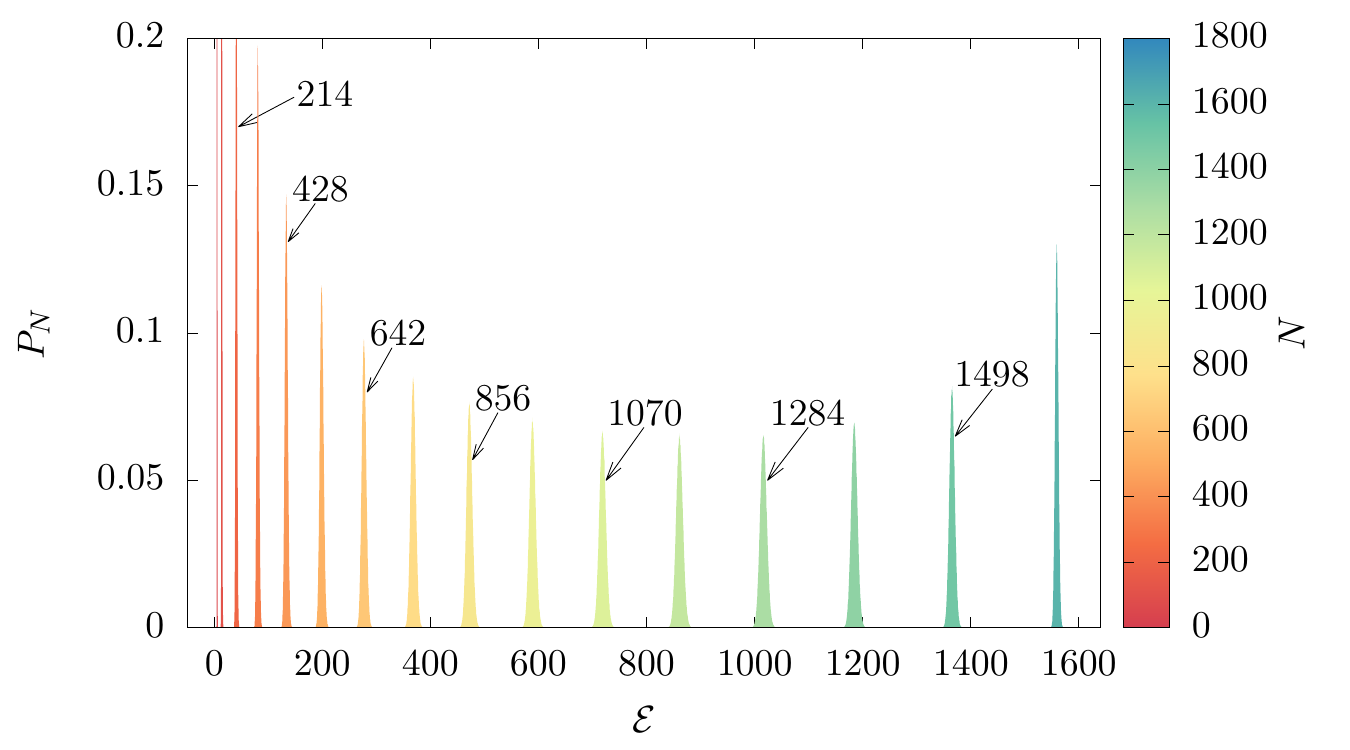}
		\caption{Probability $P_N(\mathcal{E})$ of the norm of the mapping $\mathcal{E}$ for AKE calculated at various degrees of CG'ing $N$, as obtained from a random sampling of the mapping space $\mathcal{M}$. Arrows indicate the values of $N$ for which a reconstruction of the density of states $\Omega_N(\mathcal{E})$ through the Wang-Landau algorithm has been performed.\label{fig:histo_ran}}
\end{figure}

\begin{figure}
		\includegraphics[width=\columnwidth]{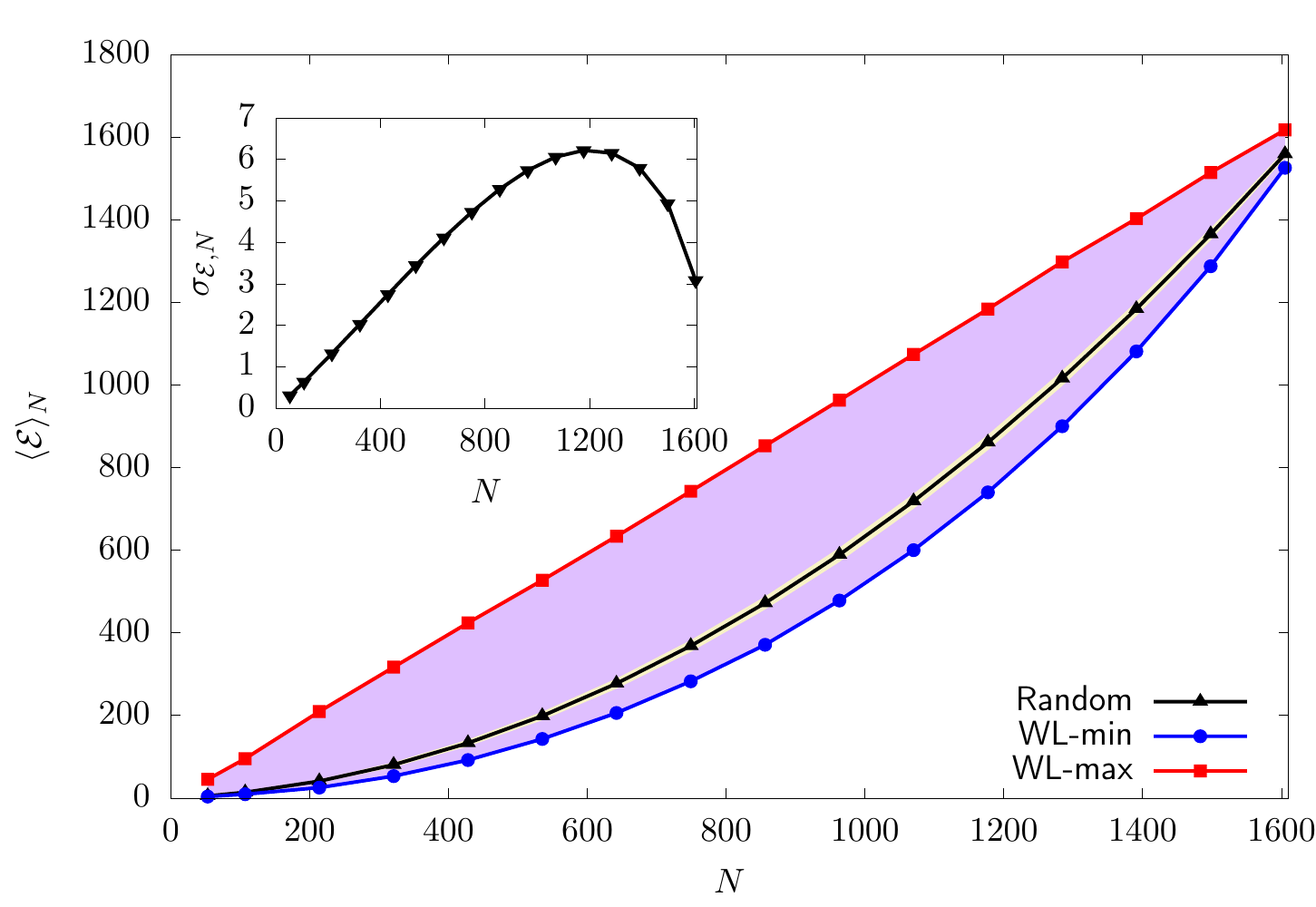}
		\caption{Inset: Standard deviation $\sigma_{\mathcal{E},N}$ of the mapping norm $\mathcal{E}$ as a function of the degree of CG'ing $N$ obtained from a random sampling of the mapping space $\mathcal{M}$ of AKE. Main plot: $N$-dependence of the average squared norm $\langle\mathcal{E}\rangle_N$ (``Random'', black line) and associated $3\sigma_{\mathcal{E},N}$ confidence interval (khaki area) as obtained from a random sampling of the mapping space of AKE, superimposed to the region covered by the set of single-window, preliminary WL runs (purple area). The minimum (``WL-min'', blue line) and maximum (``WL-max'', red line) squared norms reached by the preliminary runs are highlighted. ``WL-max'' also corresponds to the scaling $\mathcal{E}\approx N$ obtained in the case of inhomogeneous, globular mappings.\label{fig:scaling_ran_WL}}
\end{figure}

To overcome this combinatorial challenge, for each degree of CG'ing we generated $\widetilde{\Omega}_{tot}=2\cdot 10^6$ uniformly distributed random mappings as strings $\chi_i,~i=1,...,n$ of zeros and ones compatible with Eq.~\ref{eq:normalization}, and calculated the associated squared norm $\mathcal{E}$. Results for each $N$ were then binned along the $\mathcal{E}$ axis in intervals of $\delta\mathcal{E}=0.1$, and the corresponding $P_N(\mathcal{E})$ was estimated as
\begin{equation}
\label{eq:Probrand}
P_N(\mathcal{E})=\frac{1}{\delta\mathcal{E}}\frac{\widetilde{\Omega}_N(\mathcal{E})}{\widetilde{\Omega}_{tot}},
\end{equation}
where $\widetilde{\Omega}_N(\mathcal{E})$ is the number of sampled mappings with squared norm falling between $\mathcal{E}$ and $\mathcal{E}+\delta\mathcal{E}$. Note that in this way we are approximately treating as continuous the intrinsically discrete, unevenly spaced spectrum of possible norms, and the density $P_N(\mathcal{E})$---and consequently $\Omega_N(\mathcal{E})$---as piecewise constant. In this ``continuous'' limit, the normalization condition of $P_N(\mathcal{E})$ becomes
\begin{equation}
1=\sum_{\mathcal{E}}P_N(\mathcal{E})\delta{\mathcal{E}}\simeq \int d\mathcal{E}P_N(\mathcal{E}).
\end{equation}
The set of distributions $P_N(\mathcal{E})$ obtained from our random sampling of the mapping space of AKE are displayed in Fig.~\ref{fig:histo_ran}. We observe that, for each value of the CG resolution $N$, $P_N(\mathcal{E})$ is unimodal and narrowly peaked around its average squared norm,
\begin{equation}
\label{eq:norm_from_dist}
\langle\mathcal{E}\rangle_N=\int d\mathcal{E}P_N(\mathcal{E})\mathcal{E},
\end{equation}
$\langle\mathcal{E}\rangle_N$ being an increasing function of $N$. On the other hand, the standard deviation $\sigma_{\mathcal{E},N}$,
\begin{equation}
\label{eq:sig_from_dist}
\sigma_{\mathcal{E},N}=\left(\int d\mathcal{E}P_N(\mathcal{E})(\mathcal{E}-\langle\mathcal{E}\rangle_N)^2\right)^{\frac{1}{2}},
\end{equation}
is non-monotonic in the degree of CG'ing: starting from extremely small values in the case of few retained atoms (e.g. $N=53,107$ and $214$), $\sigma_{\mathcal{E},N}$ increases roughly up to $N\approx 3n/4$ and then starts to decrease, reaching zero for $N=n$---in this case only one possible mapping exists, namely the atomistic one. These features are further highlighted in  Table~\ref{tab:RSvsSP} and Fig.~\ref{fig:scaling_ran_WL}, in which we report the dependence of $\langle\mathcal{E}\rangle_N$ and $\sigma_{\mathcal{E},N}$ on the degree of CG'ing $N$ as obtained from the distributions $P_N(\mathcal{E})$ in Fig.~\ref{fig:histo_ran}. 

$\langle\mathcal{E}\rangle_N$ quantifies the average spatial homogeneity of the ensemble of CG representations that can be randomly assigned to AKE at a specific resolution. As previously discussed, maximally inhomogenous mappings, in which a chiseled chunk of the biomolecule is treated atomistically while the remainder is almost neglected, are characterised by $\mathcal{E}\approx N$. Critically, Fig.~\ref{fig:scaling_ran_WL} displays that such linear scaling lies always above the average $\langle\mathcal{E}\rangle_N$ for all degrees of coarse-graining investigated. The deviation between the two curves is non-monotonic, with a maximum obtained for $N=n/2$, and only vanishes for $N\rightarrow n$, where mappings become very dense as they collapse towards the atomistic representation. As a consequence, the CG representations one encounters by randomly probing the mapping space $\mathcal{M}$ tend to be ``sparse'' rather than compact. Furthermore, the difference between the squared norm of the globular case and $\langle\mathcal{E}\rangle_N$ is always (but for $N\approx n$) one or two orders of magnitudes larger than the standard deviation of the corresponding $P_N(\mathcal{E})$, see Fig.~\ref{fig:scaling_ran_WL}. It follows that inhomogeneous mappings lie extremely far away in the right tails of the distributions displayed in Fig.~\ref{fig:histo_ran}, thus constituting an exponentially vanishing subset of the space $\mathcal{M}$. 

\begin{figure*}[pt]
		\includegraphics[width=\textwidth]{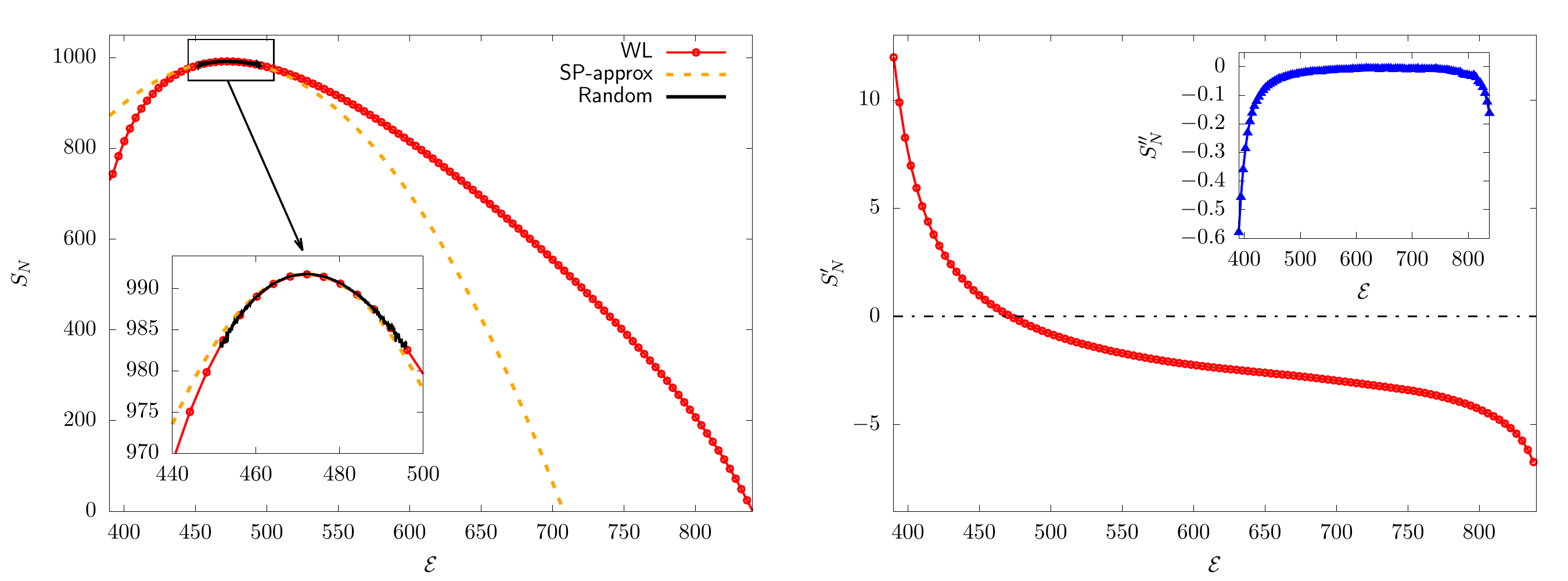}
		\caption{\emph{Left}: Logarithm of the density of states $\Omega_{N}(\mathcal{E})$ of AKE, $S_N(\mathcal{E})=\ln[\Omega_{N}(\mathcal{E})]$, for $N=856$. We report results obtained via (\emph{i}) Wang-Landau sampling (``WL'', red dotted line), vertically shifting the data so that the minimum of $S_N$ over the range of investigated norms is zero;  (\emph{ii}) a saddle-point approximation of the WL predictions (``SP-approx'', orange dashed line); and (\emph{iii}) a random drawing of CG representations (``Random'', black line), in this latter case shifting the curve so that its maximum coincides with the one the WL profile. \emph{Right}: First (main plot) and second (inset) derivatives $S'_N(\mathcal{E})$ and  $S''_N(\mathcal{E})$ of the entropy $S_N(\mathcal{E})$ determined via WL sampling for $N=856$. \label{fig:entropy_856}}
\end{figure*}

The suppression of the statistical weight associated to high-norm, globular CG representations of AKE in the space of all possible ones is not surprising, and is solely driven by entropic effects. Indeed, at least for small and intermediate $N$, it is extremely unlikely that a completely random selection of retained atoms across the biomolecule will result in their dense confinement within sharply-defined spatial domains of the system, just as it is unlikely for a gas to occupy only a small fraction of the volume in which it is enclosed. Interestingly, this latter analogy can be pushed further by noting that the squared norm $\mathcal{E}(M)$, see Eq.~\ref{eq:rescaled_norm} and~\ref{eq:norm_gauss}, is akin to the negative configurational energy of a lattice gas living on the irregular grid defined by the protein's conformation, whose particle interact via a hard-core, short-range potential followed by an attractive Gaussian tail. In this context, the selection operators $\chi_{M,i}=0,1,~i=1,...,n$ of a mapping $M$ with $N$ retained atoms can be interpreted as the set of occupation numbers describing a distribution of the $N$ particles of the gas on the $n$ available lattice sites. It follows that compact CG representations of AKE, located in the large-$\mathcal{E}$ limit of $P_N(\mathcal{E})$, are just as challenging to randomly sample within the space $\mathcal{M}$ as are the low-energy configurations of the gas in which the $N$ particles spontaneously occupy only a fraction of the available volume. The implications of this analogy will be thoroughly explored in Sec.~\ref{sec:phasetrans}.

The strongly entropy-driven distribution of mappings calls for the introduction of enhanced sampling techniques to boost the exploration of the mapping space; in this work, we resort to the algorithm proposed by Wang and Landau (WL) \cite{wang2001determining,wang2001efficient,shell2002generalization,barash2017control}. For each CG resolution $N$, the aim is to obtain a \emph{uniform} sampling of the possible mapping norms $\mathcal{E}$ across the space $\mathcal{M}$, in contrast to the set of narrowly-peaked probability distributions displayed in Fig.~\ref{fig:histo_ran}. In principle, this is attained by setting up a Markov chain Monte Carlo simulation in which a transition between two subsequent mappings $M$ and $M'$---both retaining $N$ atoms---is accepted with probability $\alpha$ given by \cite{wang2001determining}
\begin{eqnarray}
\label{eq:WL_acceptance}
&&\alpha_{M\rightarrow M'}=\text{min}\left[1,\frac{\Omega_N(\mathcal{E}(M))}{\Omega_N(\mathcal{E}(M'))}\right]\nonumber\\
&&=\text{min}\left[1,\exp{\left(-[S(\mathcal{E}(M'))-S(\mathcal{E}(M))]\right)}\right],
\end{eqnarray}
where $\Omega_N(\mathcal{E})$ is the density of states defined in Eq.~\ref{eq:pgiven} while $S_N(\mathcal{E})=\ln[\Omega_N(\mathcal{E})]$ is the corresponding microcanonical entropy. 

When compounded with a symmetric proposal probability $\pi$ for the attempted move, $\pi_{M\rightarrow M'}=\pi_{M'\rightarrow M}$, the Markov chain in Eq.~\ref{eq:WL_acceptance} would generate, after an initial relaxation transient, CG representations distributed according to $p(M)\sim 1/\Omega_N(\mathcal{E}(M))$ \cite{wang2001determining}, resulting in a flat histogram $P_N(\mathcal{E})$ of visited norms \emph{over the whole range of possible ones} \cite{barash2017control}. 

In practice, however, the density of states in Eq.~\ref{eq:WL_acceptance} is not known a priori. The power of WL approach resides in its ability to self-consistently obtain $\Omega_N(\mathcal{E})$ through a sequence $k=1,...,K$ of non-equilibrium simulations in which increasingly accurate approximations $\bar{\Omega}^k_N(\mathcal{E})$ to the exact result are generated, iterations being stopped when the desired precision is achieved \cite{wang2001determining,wang2001efficient}. For the sake of brevity, we here omit an exhaustive discussion of the general algorithmic workflow behind WL sampling as well as an in-depth description of the specific implementation employed in this work; these details are provided in Appendix~\ref{app:wangland}.

In the WL reconstruction of a density of states such as $\Omega_N(\mathcal{E})$, knowledge of the sampling boundaries proves extremely beneficial to the accuracy and rate of convergence of the self-consistent scheme \cite{wust2008hp}. For each degree of CG'ing investigated, we thus initially performed a preliminary, non-iterative WL run to approximately locate the minimum and maximum mapping norms $\mathcal{E}_{min}(N)$ and $\mathcal{E}_{max}(N)$ achievable for AKE at that specific CG resolution, and consequently bound the support of the corresponding $\Omega_N(\mathcal{E})$. 

The results for $\mathcal{E}_{min}(N)$ and $\mathcal{E}_{max}(N)$ obtained from this analysis are presented in Fig.~\ref{fig:scaling_ran_WL} and Table~\ref{tab:WLwindef} of Appendix~\ref{app:wangland}. We observe that the mapping norms visited by the set of preliminary WL runs extend, for all values of $N$, over a significantly wider range compared to the one obtained by random sampling. Remarkably, the maximum norm $\mathcal{E}_{max}(N)$ exhibits a linear dependence on $N$ that is fully compatible with the one associated to globular CG representations, $\mathcal{E}_{max}(N)\approx N$, highlighting that the WL approach succeeds in exploring this entropically suppressed region of the mapping space. Furthermore, Fig.~\ref{fig:scaling_ran_WL} displays that the minimum norm $\mathcal{E}_{min}(N)$ identified by the preliminary runs lies always below the average $\langle\mathcal{E}\rangle_N$ for all values of $N$. In contrast to globular mappings, CG representations living in this low $\mathcal{E}$ limit are \emph{maximally homogeneous}, that is, retained atoms are scattered throughout the molecular structure as uniformly as possible. This class constitutes another exponentially vanishing subset of the mapping space: in the gas picture, it would correspond to the ensemble of configurations in which gas particles are \emph{regularly} distributed within the available volume.

Having approximately identified the range of mapping norms achievable for AKE at each CG resolution, we subsequently moved to the determination of the associated densities of states $\Omega_N(\mathcal{E})$ via the iterative WL scheme, see Appendix~\ref{app:wangland} for all technical details. Calculations were only performed for a subset of degrees of CG'ing, namely those in which the number of retained atoms $N$ is an integer multiple of the number of residues composing the biomolecule, $N=i\cdot 214,~i=1,...,7$.

To speed-up convergence of the algorithm, for each $N$ we slightly reduced the range of norms $[\mathcal{E}_{min},\mathcal{E}_{max}]$ with respect to the one predicted by the explorative WL runs, see Table~\ref{tab:WLwindef} in Appendix~\ref{app:wangland}. This interval was then divided into a set of overlapping windows in which independent WL simulations were performed \cite{wang2001efficient}. The resulting partial densities of states were a posteriori combined to determine the cumulative $\Omega_N(\mathcal{E})$ up to a global multiplicative factor, or, in our case, the entropy $S_N(\mathcal{E})=\ln[\Omega_N(\mathcal{E})]$ up to an additive constant.

WL estimates of the entropy $S_N(\mathcal{E})$ are presented in Fig.~\ref{fig:entropy_856} for $N=856$, while results for all the other degrees of CG'ing are reported in Fig.~\ref{fig:dos_N_WL} of Appendix~\ref{app:wangland}. In all cases, we observe that the behavior of $S_N$ is non-monotonic in $\mathcal{E}$, exhibiting a unique maximum as the mapping norm moves from the left to right boundary of the range of investigated ones---that is, in transitioning from extremely homogeneous to maximally globular CG representations. As $\Omega_N(\mathcal{E})=\exp[S_N(\mathcal{E})]$, this result confirms how these two limiting classes of mappings constitute regions of exponentially vanishing size within the broad space $\mathcal{M}$. At the same time, the overall shape of $S_N$ strongly depends on the degree of CG'ing: while for high $N$ entropy profiles are nearly symmetric around their maximum, they become increasingly skewed as fewer and fewer atoms are employed to represent the macromolecule. This asymmetry becomes apparent by performing, for each CG resolution, a quadratic expansion of $S_N$ around its maximum, 
\begin{equation}
S_N(\mathcal{E})\simeq S_N(\tilde{\mathcal{E}}(N))+\frac{1}{2}S''_N(\tilde{\mathcal{E}}(N))(\mathcal{E}-\tilde{\mathcal{E}}(N))^2,
\label{eq:quadrapprox}
\end{equation}
where $\tilde{\mathcal{E}}(N)$ is the norm at which the first derivative $S'_N$ of the entropy vanishes, and $S''_N(\tilde{\mathcal{E}}(N))$ is the corresponding second derivative---the dependence of $S'_N$ and $S''_N$ on $\mathcal{E}$ being displayed in Fig.~\ref{fig:entropy_856} for $N=856$. The accuracy of this parabolic, symmetric approximation in reproducing the exact $S_N$ over the whole $\mathcal{E}$-range increases with the number of retained atoms, see Fig.~\ref{fig:entropy_856} and~\ref{fig:dos_N_WL}, especially as far as the limit of high mapping norms is concerned.

\begin{figure*}[ht] 

\centering
    \begin{subfigure}[b]{\columnwidth}
        \includegraphics[width=\columnwidth]{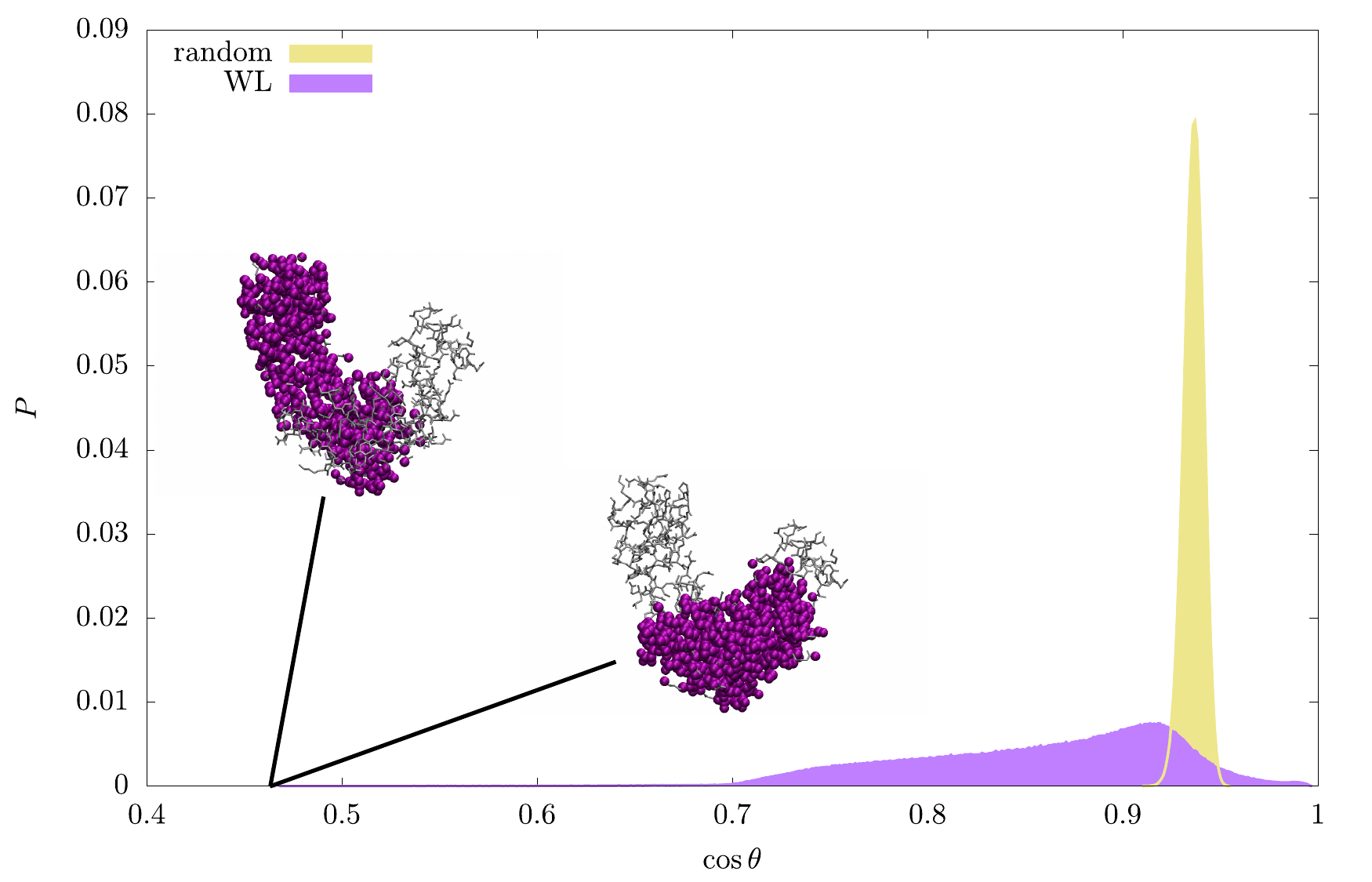}
    \end{subfigure} %
    \begin{subfigure}[b]{\columnwidth}    
        \includegraphics[width=\columnwidth]{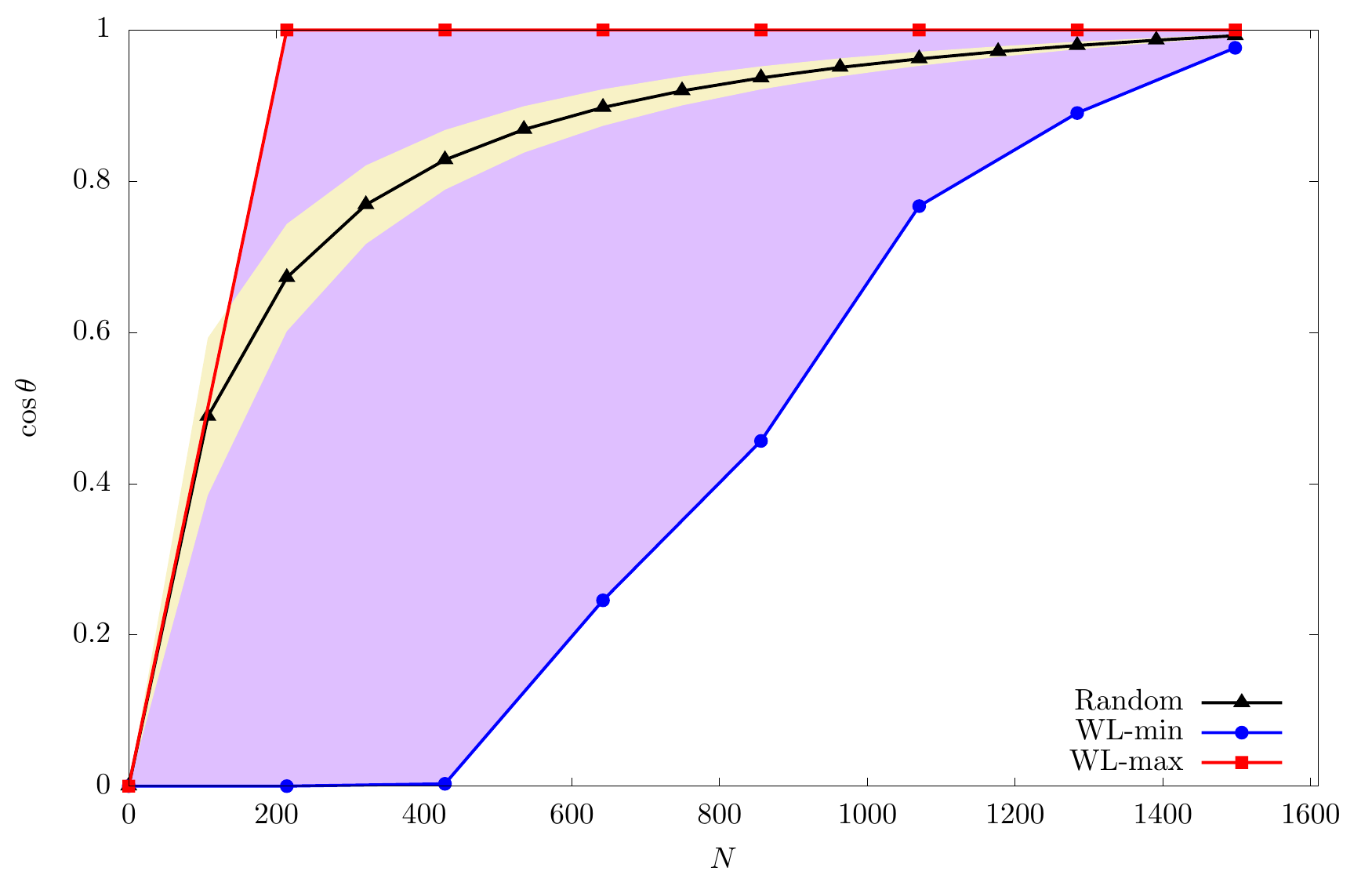}
    \end{subfigure} 
    \caption{\emph{Left}: histogram of cosine values extracted from random CG mappings (yellow) and WL CG mappings (purple, see main text) for AKE with $N=856$ sites. Elements of $\mathcal{M}$ with the lowest value of the cosine ($\cos \theta = 0.457$) are shown; such value corresponds to an angle of $63.25$ degrees. \emph{Right}: range of cosine values covered by the two data sets when $N$ is changed. The dotted black line shows the average value of $\cos \theta$ over the different random data sets and the yellow region represents the points within $3 \sigma$ from the mean. The red (blue) dotted lines report the maximum (minimum) values of $\cos  \theta$ inside WL data sets, respectively.}
     \label{fig:histo_cosines}
\end{figure*}

Finally, it is interesting to test the predictions of WL sampling against the results obtained via a completely random exploration of the mapping space. To this end, Fig.~\ref{fig:entropy_856} and Fig.~\ref{fig:dos_N_WL} include a comparison between the WL entropies $S_N$ and their random counterparts $S^{ran}_N$, the latter defined as $S^{ran}_N(\mathcal{E})=\ln [P_N(\mathcal{E})]+C_N$, where $P_N(\mathcal{E})$ are the probability densities presented in Fig.~\ref{fig:histo_ran} and the constants $C_N$ are set so that the maxima of $S^{ran}_N$ and $S_N$ coincide. For each value of $N$ the two profiles are in perfect agreement, thus confirming the accuracy of the self-consistent WL scheme in determining the density of states of a system. Critically, results for $S^{ran}_N$ only extend over a very narrow range of mapping norms, centered around the value $\tilde{\mathcal{E}}(N)$ for which the maximum of the entropy is attained. It is therefore largely unfeasible, by randomly drawing CG representations, to exhaustively explore the mapping space $\mathcal{M}$ of a macromolecule. 

To provide a more quantitative measure of the consistency between random and WL sampling results, for each degree of CG'ing we recalculated the average and variance of the mapping norm, see Eqs.~\ref{eq:norm_from_dist} and \ref{eq:sig_from_dist}, starting from the WL entropies $S_N$. These are used to compute $P_N(\mathcal{E})$ making use of a saddle-point approximation of Eq.~\ref{eq:pgiven}, namely
\begin{eqnarray}
\label{eq:sad_point}
P_N(\mathcal{E})=\frac{\Omega_N(\mathcal{E})}{\Omega_N}=\frac{\exp[S_N(\mathcal{E})]}{\int d\mathcal{E}\exp[S_N(\mathcal{E})]}\simeq\nonumber \\
\hspace*{-0.7cm}\left(\frac{|S''_N(\tilde{\mathcal{E}}(N))|}{2\pi}\right)^{\frac{1}{2}}\exp\left[{\frac{1}{2}S''_N(\tilde{\mathcal{E}}(N))(\mathcal{E}-\tilde{\mathcal{E}}(N))^2}\right],
\end{eqnarray}
where in the last step of Eq.~\ref{eq:sad_point} we made use of the quadratic expansion of $S_N$ defined in Eq.~\ref{eq:quadrapprox}. Within the saddle point approximation, one has $\langle\mathcal{E}\rangle_N=\tilde{\mathcal{E}}(N)$, $\tilde{\mathcal{E}}(N)$ being the position of the maximum of $S_N$, and $\sigma_{\mathcal{E},N}=|S''_N(\tilde{\mathcal{E}}(N))|^{-\frac{1}{2}}$: these predictions are found to be in perfect agreement with their random sampling counterparts, results being presented in Table~\ref{tab:RSvsSP}.

\subsection{Inner product distributions}
\label{sec:cosine}

Here we proceed to the description of the mapping space $\mathcal{M}$ from the perspective of the inner product between its elements. Following the same scheme of Sec.~\ref{sec:norm}, we here focus on the cosine between mappings that are constrained to share the same resolution $N$. To fulfil this purpose we can compute the probability $P_{NN}(\cos \theta)$ of observing a value of $\cos \theta$ provided that this constraint is satisfied:
\begin{eqnarray}
\label{eq:}
P_{NN}(\cos \theta) = \frac{\Omega_{NN}(\cos \theta)}{\Omega^2_{N}},
\end{eqnarray}
that is, the ratio between the number of mapping pairs whose cosine is equal to $\cos \theta$, $\Omega^2_{NN}(\cos\theta)$, and the total number of possible pairs $\Omega^2_{N}$. We can now investigate how the average \textit{degree of parallelism} between two mappings changes when considering randomly selected mappings or more peculiar elements of $\mathcal{M}$.

In this section we compare two data sets, each one containing $10^6$ elements: the first is obtained by computing the cosine between two mappings in which the retained sites have been picked randomly; the second data set is constructed in a more sophisticated manner, making use of the WL sampling scheme to collect mappings that uniformly span the range of accessible values of $\mathcal{E}$, which is known from the previous section. More specifically, we start a WL exploration as in Sec.~\ref{sec:norm} over this range and, when all the reference bins have been visited at least once, we start saving a mapping every $1656$ Monte Carlo moves. Mappings are saved in different macro-bins, each one covering an interval of amplitude $20$ (in terms of units of $\mathcal{E}$). Sampling ends when 5000 mappings are saved in each box, without considering the convergence of the WL algorithm. The data set is then generated by computing the cosine (Eq.~\ref{eq:cosine}) between randomly selected pairs of mappings extracted through this procedure. Importantly, the WL sampling scheme produces a pool of potentially correlated mappings and the chance of collecting similar elements of $\mathcal{M}$ cannot be excluded.

Fig.~\ref{fig:histo_cosines} (a) shows the histograms between the two data sets for $N=856$. While the random cosine distribution displays a narrow peak around its average value, the WL histogram is more distributed, reflecting the increased diversity of the data set. Indeed, the latter histogram spans values that range from $\sim 1$, obtained when two mappings are perfectly parallel, to $0.457$, when two mappings are as orthogonal as possible given the constraints of the lattice, that is, the protein structure. In Fig.~\ref{fig:histo_cosines} (a) we also report a graphical rendering of the two maximally orthogonal mappings, which possess a high value of $\mathcal{E}$ ($\mathcal{E}=847.32$ and $\mathcal{E}=843.82$, respectively) and cover different regions of the enzyme's structure.

In Fig.~\ref{fig:histo_cosines} (b) we extend these considerations to different values of $N$, namely those employed in Sec.~\ref{sec:norm}. The random distribution is always confined in a narrow interval of values of $cos \theta$, while WL data sets are capable of spanning a much wider range. In particular, for sufficiently small values of $N$, it is possible to retrieve maximally parallel ($\cos \theta = 1$) and maximally orthogonal ($\cos \theta = 0$) mappings inside the WL dataset. This is made possible by the fact that, at such low values of $N$, it is possible to confine retained sites in two separate regions of the protein structure.

\section{Lattice gas analogy and phase transitions}
\label{sec:phasetrans}

As anticipated in Sec.~\ref{sec:explore}, the reduced representation discussed in the present work, in which a mapping is defined in terms of a {\it decimation} of the atoms available on the molecular structure, suggests the analogy with a lattice gas. Also in this case, in fact, we have a number $n$ of nodes that can be occupied by $N \leq n$ sites, each node being accessible to a single site at a time--thus implementing a hard-core repulsion.

The role of the energy can be played by the norm of the mapping. In analogy with a lattice gas, we expect that if two retained sites are close to each other, they feel an attractive interaction, thereby reducing the energy. The total energy of the system can be written as:
\begin{eqnarray}\label{pt:1}
E(M) = - \mathcal E(M).
\end{eqnarray}

In the previous sections we have obtained the density of states in terms of the mapping norm, $\Omega_N = \Omega_N(\mathcal E)$. Making use of Eq.~\ref{pt:1} we can thus write:
\begin{eqnarray}\label{pt:2}
\Omega_N(E) = \Omega_N(- \mathcal E).
\end{eqnarray}

Let us now consider a system governed by the lattice Hamiltonian in Eq.~\ref{pt:1} at equilibrium with a reservoir at temperature $T = \beta^{-1}$. The partition function of such system can be expressed in terms of $\Omega_N(E)$ {\it via}:
\begin{eqnarray}\label{pt:3}
\mathcal Z_N(\beta) &=& \int dE\ e^{-\beta E}\Omega_N(E)\\ \nonumber
&\equiv& \int dE\ e^{- (\beta E - S_N(E))}
\end{eqnarray}
where we used the relation $S_N(E) = \ln \Omega_N(E)$ to define the entropy. Eq.~\ref{pt:3} enables us to compute the dimensionless Helmholtz free energy as:
\begin{eqnarray}\label{pt:4}
\beta F_N(\beta) &=& - \ln \mathcal Z_N(\beta)\\ \nonumber
&=& - \ln \int dE\ e^{- (\beta E - S_N(E))}.
\end{eqnarray}

While the logarithm of the integral can be theoretically and numerically cumbersome to compute, it is possible to obtain a reasonable estimate of $\beta F_N$ through a saddle point approximation. Specifically, we can expect that the integral is approximately equal to the largest integrand, so that:
\begin{eqnarray}\label{pt:5}
\int dE\ e^{- (\beta E - S_N(E))} \simeq C\ \max_E \left(e^{- (\beta E - S_N(E))}\right)
\end{eqnarray}
where $C$ is an immaterial constant. This approximation provides us with a definition of the free energy that is equivalent to the Legendre-Fenchel transform:
\begin{eqnarray}\label{pt:6}
\beta F_N(\beta) \simeq \min_E \left(\beta E - S_N(E)\right).
\end{eqnarray}

The thermodynamics of the lattice gas at thermal equilibrium can thus be retrieved computing Eq.~\ref{pt:6} for a given value of $N$ at all values of $\beta$.

\begin{figure}[pt]
		\includegraphics[width=\columnwidth]{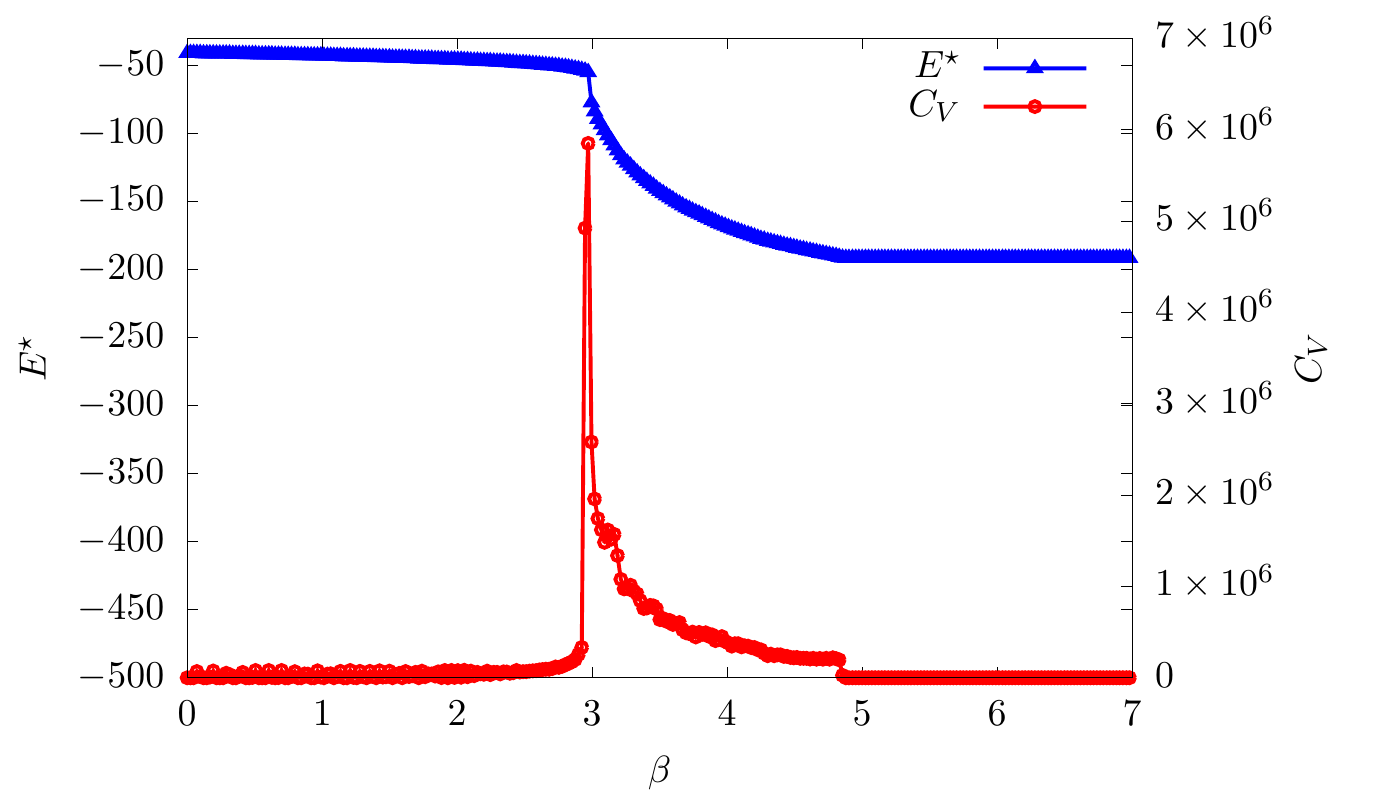}
		\caption{Heat capacity $C_V$ (red circles, right ordinate) and value of the energy $E^\star$ corresponding to the minimum of the free energy (blue triangles, left ordinate) as functions of the inverse temperature $\beta$ for the system with $N = 214$. $E^\star$ decreases monotonically with $\beta$, indicating that higher temperatures correspond to higher values of the average internal energy of the lattice gas, as expected; however, a jump discontinuity in $E^\star$ appears in correspondence of the same value $\beta_{gl}$ for which the heat capacity features a sharp peak, suggesting the occurrence of a first order phase transition that separates two distinct phases: a gas (low $\beta$) from a liquid (high $\beta$) for the lattice gas model, and, correspondingly, a sparse phase from a dense, localised phase in the case of mappings.\label{fig:ebeta}}
\end{figure}

It is particularly instructive to investigate the temperature dependence of $E^\star$, defined as the value of the energy for which $\beta E - S(E)$ reaches its minimum. In Fig.~\ref{fig:ebeta} (blue curve, left ordinate) we report this function for $N = 214$: it is possible to observe that $E^\star = E^\star(\beta)$ decreases monotonically, i.e., the lower the temperature, the lower the value of the energy--which corresponds to higher values of the mapping norm. At a particular value $\beta_{gl}$ of the inverse temperature, however, $E^\star$ drops abruptly: in this context, such behaviour is suggestive of a first-order, discontinuous phase transition.

\begin{figure}
		\includegraphics[width=\columnwidth]{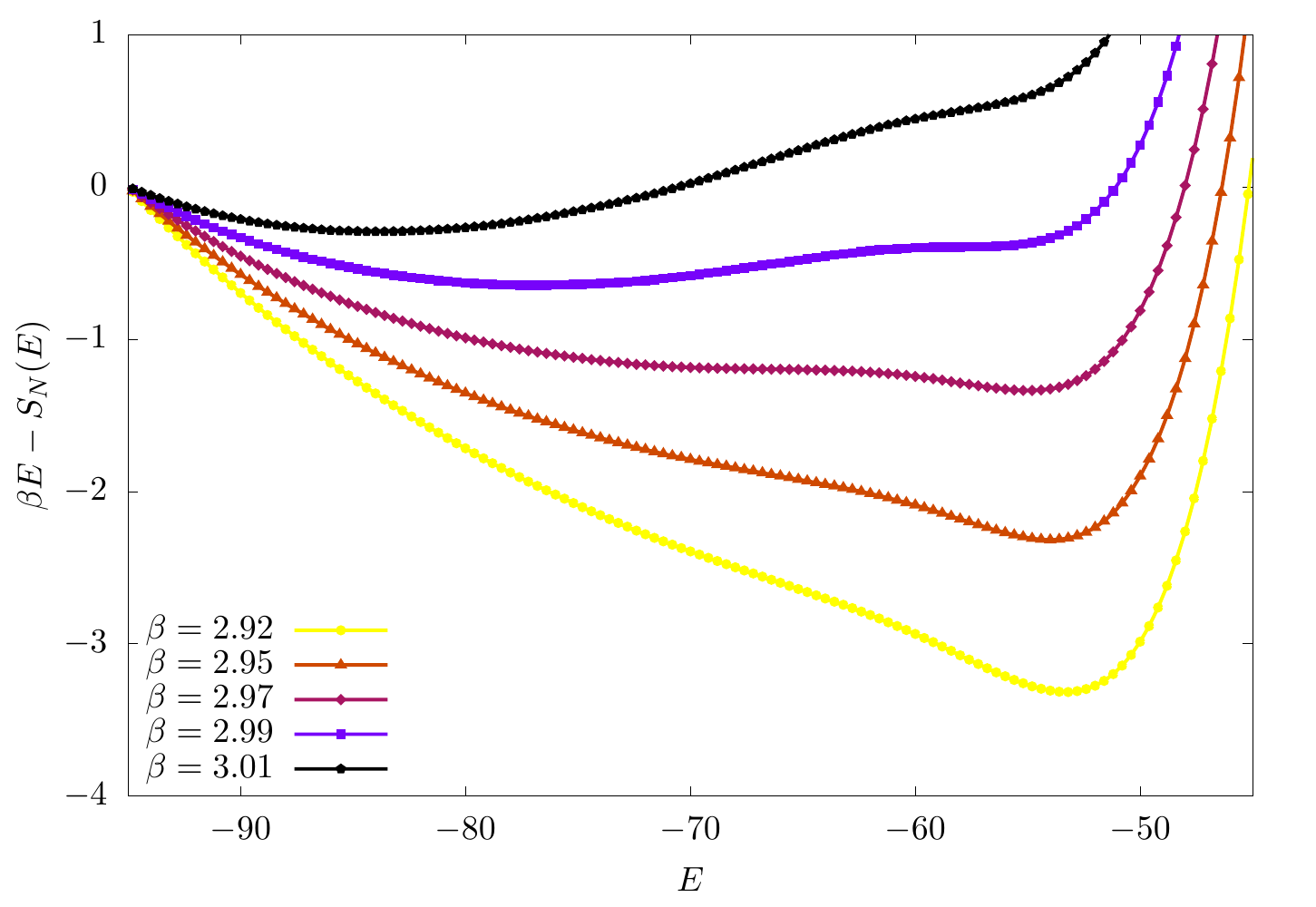}
		\caption{
		The Helmholtz free energy $\beta F$ as a function of the energy for different values of the inverse temperature $\beta$. The curves have, in general, a unique absolute minimum; however, as $\beta$ increases, a metastable minimum appears that, for a particular value of the inverse temperature, becomes degenerate. The presence of a small but appreciable barrier between the two minima makes the position of the absolute minimum, $E^\star$, shift abruptly from one to the other, as can be seen in Fig. \ref{fig:spec_heat}, thus making $E^\star(\beta)$ discontinuous.
		\label{fig:betaF}}
\end{figure}

To gain further insight, we have computed the shapes of $\beta E - S(E)$ for values before and after $\beta_{gl}$. These functions, reported in Fig.~\ref{fig:betaF}, indeed show two minima separated by a relatively low barrier; increasing $\beta$, the absolute minimum shifts from the right to the left, crossing a point for which the two are essentially degenerate. This is the point of coexistence of two distinct ``phases'' of our lattice gas: a low density one corresponding to distributed mappings (high energy), and one ascribable to more dense, compact conglomerates of sites (low energy). The critical nature of the transition from one regime to the other is confirmed by the inspection of the heat capacity, computed as
\begin{eqnarray}\label{pt:7}
C_V = - \beta^2 \frac{\partial^2 (\beta F)}{\partial \beta^2}
\end{eqnarray}
and reported in Fig.~\ref{fig:ebeta} (red curve, right ordinate). The sharp, asymmetric peak in $C_V$, located at the value $\beta_{gl}$ of the inverse temperature, shows that the lattice gas crosses a phase transition between a gas and a liquid phase.

A crucial role in this behaviour is played by the number of coarse-grained sites. In fact, as this number increases, the system acquires the possibility of crossing a second phase transition: for example, in the case of $N = 1070$, besides the gas-liquid one, it is possible to observe a second, even sharper discontinuity in $E^\star$ for a value of the inverse temperature $\beta_{ls} > \beta_{gl}$. This temperature separates the liquid from the solid phase: when the lattice gas particles are sufficiently many, and the temperature sufficiently low, the system can ``freeze'' in particularly dense mappings with very low entropy. Also in this case, the inspection of the heat capacity (Fig.~\ref{fig:spec_heat} in Appendix~\ref{app:latticegas}) supports the interpretation of this as a phase transition. Finally, if the number of sites is too large (e.g. $N = 1498$) no transition is observed, see Fig.~\ref{fig:spec_heat}.

The observations reported in this section resonate with those made by Foley and collaborators in a recent work \cite{foley2020exploring}: there, they observed a phase transition in a system whose degrees of freedom were the retained sites of a reduced model of proteins. In that case, the energy of a given mapping was obtained from the calculation of the spectral quality of the associated model, a quantity related to the sum of the eigenvalues of the covariance matrix obtained integrating exactly a Gaussian network model (GNM). While apparently very distinct, the spectral quality and the norm of the mapping might bear substantial similarities: in fact, the former entails information about a very simple model, whose mechanical and thermodynamical properties are completely determined by the contact matrix of the underlying protein structure. It is thus reasonable to guess that the mapping norm provides, in an effective, efficient, and transparent manner, information akin to that entailed in the spectral quality about the sparsity or localisation of the retained sites in a given mapping. If and up to which degree these two quantities are related, and how intimately this relation depends on the Gaussian nature of the GNM, requires further investigations that will be the object of future studies.

In conclusion of this section we note that the phase transition separates mappings so diverse that they can be associated to qualitatively different phases. It is thus natural to wonder if and how these phases are organised in the metric space induced by the norm of the mapping, and what information the exploration of the latter can bring about the system it is applied to. To provide an answer to these questions, the next section is devoted to the topological characterisation of the mapping space.

\section{Topology}
\label{sec:topology}

In the previous sections we analyzed the mapping space $\mathcal{M}$ in terms of the mapping norm $\mathcal{E}$ and of the cosine between its constituent elements. Here, we discuss the distance $\mathcal{D}$ (Eqs.~\ref{eq:dist_dotmapp}, \ref{eq:rescaled_norm}) between members of $\mathcal{M}$ with the aim of showing, once again, that a \textit{peculiar} choice of retained CG sites, i.e. one impossible to obtain with random sampling, displays non-trivial statistical properties that reflect in the topological organization of the mapping space. 

\subsection{Topology of the mapping norm space}

Without loss of generality\footnote{The general validity of the discussion presented here is supported by the results obtained for the case $N = 856$, which are reported in Fig. \ref{fig:sketch_map_sigma25} of the Appendix.}, we restrict our investigation to the case $N=214$, which is the number of amino acids of adenylate kinase. Here we generate a data set of mappings following the protocol explained in Sec.~\ref{sec:cosine}; in this case, the range of values of $\mathcal{E}$ is narrower and only $10$ macro-bins of amplitude $20$ are explored. The data set is constructed by randomly selecting $100$ elements for each of the macro-bins, resulting in $1000$ CG mappings that homogeneously span the accessible values of $\mathcal{E}$.

\begin{figure*}
\centering
        \includegraphics[width=\linewidth]{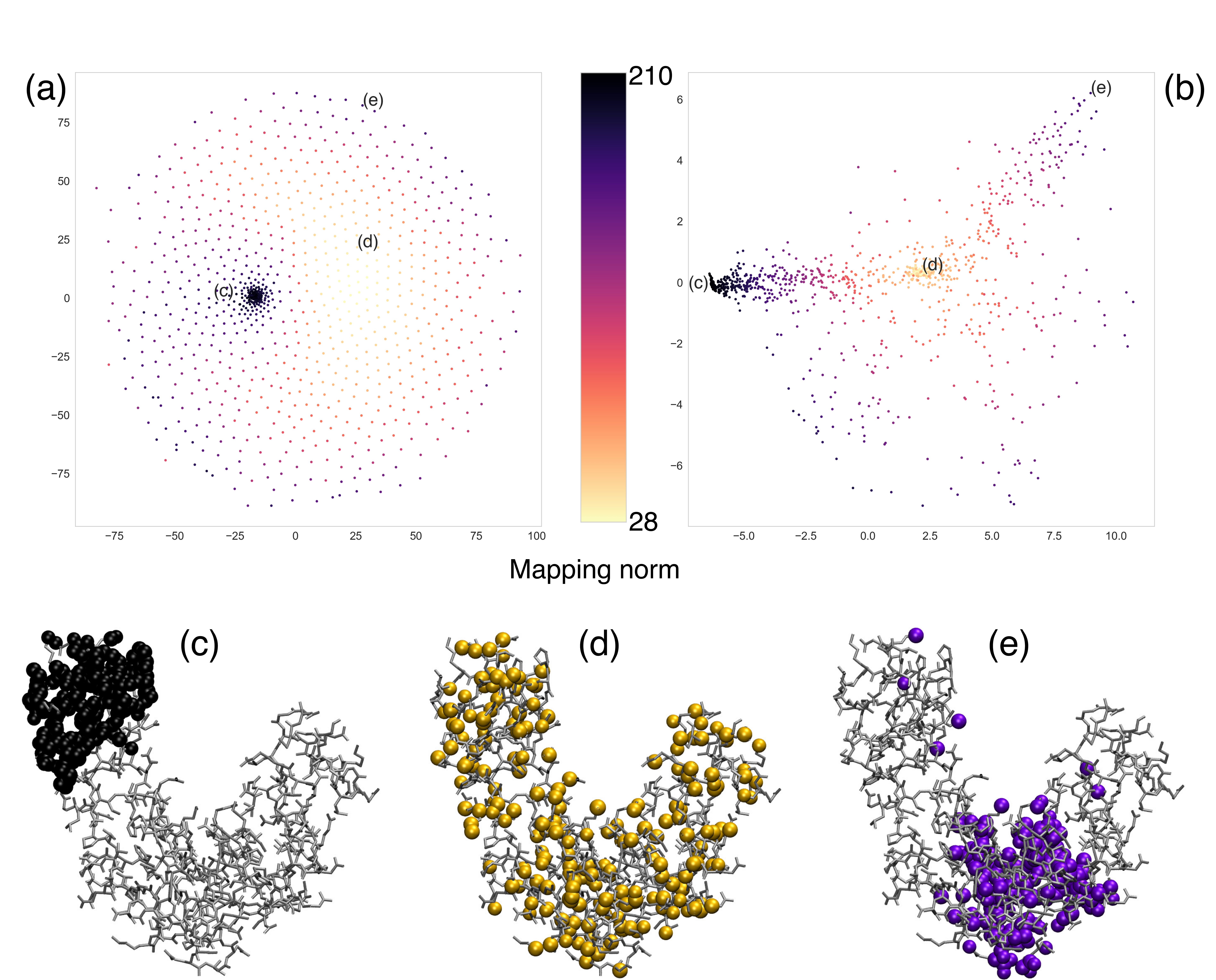}
   	\caption{\label{fig:sketch}\emph{Top}: topology of the mapping space $\mathcal{M}$ in 2D, obtained with the sketch map algorithm \cite{ceriotti2011simplifying, ceriotti2013demonstrating}. The algorithm requires six parameters, namely $\sigma_d$, $a_d$, $b_d$ in the low resolution space and $\sigma_D$, $a_D$, $b_D$ in the original, high resolution one. We select $\sigma_D = \sigma_d = 2$ for subfigure (a) and $\sigma_D = \sigma_d = 20$ for subfigure (b), while $a_d = b_d = 2$ and $a_D = b_D = 5$ in both cases. Mappings are depicted with different colors depending on their norm $\mathcal{E}$. We note that a different choice for $\sigma_D$ and $\sigma_d$ results in a completely different 2D embedding (see \cite{ceriotti2011simplifying} for a detailed explanation). \emph{Bottom}: three different mappings located in three separated regions of the plane in (a) and (b). Mappings in subfigures (c) and (e) possess very high values of $\mathcal{E}$ and are localised in different domains of the protein. It is interesting to notice that sparse mappings, such as the one in subfigure (d), are clustered in the same region in (b) but not in (a).}
\end{figure*}

The sketch map algorithm \cite{ceriotti2011simplifying, ceriotti2013demonstrating} is employed to embed $1000$ points from the high-dimensional space of mappings $\mathcal M$ into a two-dimensional plane, however preserving as faithfully as possible the relative distances among them--that is to say that nearby points in the mapping space are mapped onto nearby points on the $2$D space, see Fig.~\ref{fig:sketch}. The two critical parameters of the algorithm are $\sigma_d$ and $\sigma_D$, which modulate how \textit{far} and \textit{close} points are in the low (high) resolution space \cite{ceriotti2011simplifying}. To provide the reader with a feeling of the impact that these parameters have on the structure of the low-dimensional representation, we report the embeddings obtained for a low (Fig.~\ref{fig:sketch}\textit{a}) and high (Fig.~\ref{fig:sketch}\textit{b}) value of $\sigma_d$ and $\sigma_D$.

In the first case, reported in Fig.~\ref{fig:sketch}\textit{a} and referring to low values of the $\sigma$ parameters, data points are in general very sparse, uniformly distributed on the plane, with the exception of a few groups of points that accumulate in denser regions: these are particularly compact mappings localised in distinct regions of the molecule. In fact, as the $\sigma$ parameters are increased (thus ``squeezing" points in the low-D embedding), see Fig.~\ref{fig:sketch}\textit{b}, the points corresponding to sparse, uniform mappings are collapsed in a small region, as it happens to points representing specific denser mappings as well; points already grouped in the previous representation remain close to each other.

The high-$\sigma$ embedding highlights two relevant features: first, the presence of specific regions with qualitatively distinct mapping properties; these are either very sparse, but necessarily similar one to the other (Fig.~\ref{fig:sketch}\textit{d}), or very dense, with atoms localised in different domains of the molecule (Figs.~\ref{fig:sketch}\textit{c,e}). The distance among the latter is necessarily large, since the retained sites cover non-overlapping regions.

The second relevant feature is that different groups of points, associated to qualitatively distinct types of mappings, can be connected one to the other only ``passing through'' a third one, as in the case of mapping \textit{c} going to \textit{e} through \textit{d}. This is suggestive of the presence of {\it routes} in mapping space that join points having the same value of the norm, which however cannot be connected through ``iso-$\mathcal E$'' paths: in order to transform mappings such as that in \textit{c} into that in \textit{e} through a sequence of single-site changes (i.e. one retained atom is discarded, a formerly discarded one is now retained) one cannot but increase or decrease the value of the norm.

\subsection{Topology of mapping entropy space}

The mapping norm $\mathcal{E}$ is only one of the observables that can be exploited to investigate the topology of $\mathcal{M}$. Here, we focus on the mapping entropy $S_{map}$ \cite{Shell2008, rudzinski_2011, Shell2012, foley2015impact, giulini2020information}, which is a measure of the intrinsic information loss that is inherent to the process of dimensionality reduction operated by a mapping. While $\mathcal{E}$ depends only on the geometric properties of a single protein conformation, $S_{map}$ is calculated from an ensemble of configurations sampled according the Boltzmann distribution (see ref. \cite{giulini2020information} for the details). $S_{map} (M)$ contains more information than $\mathcal{E} (M)$, since it makes explicit use of the average structural and thermodynamical properties of the system.

Here we employ a data set of $1968$ CG mappings with $N=214$ covering a wide range of values of $S_{map}$, generated by us in a previous work \cite{errica2021deep}; the relations among these mappings are then quantified in terms of their distance $\mathcal{D}$, taking the enzyme crystal structure as a reference. With respect to this, it is worth keeping in mind that $\mathcal{D}$ intimately depends on this reference, and mappings that lie close to each other when a given structure is employed might turn out to be closer or further away from each other when a different conformation is used.

\begin{figure*}
		\includegraphics[width=\textwidth]{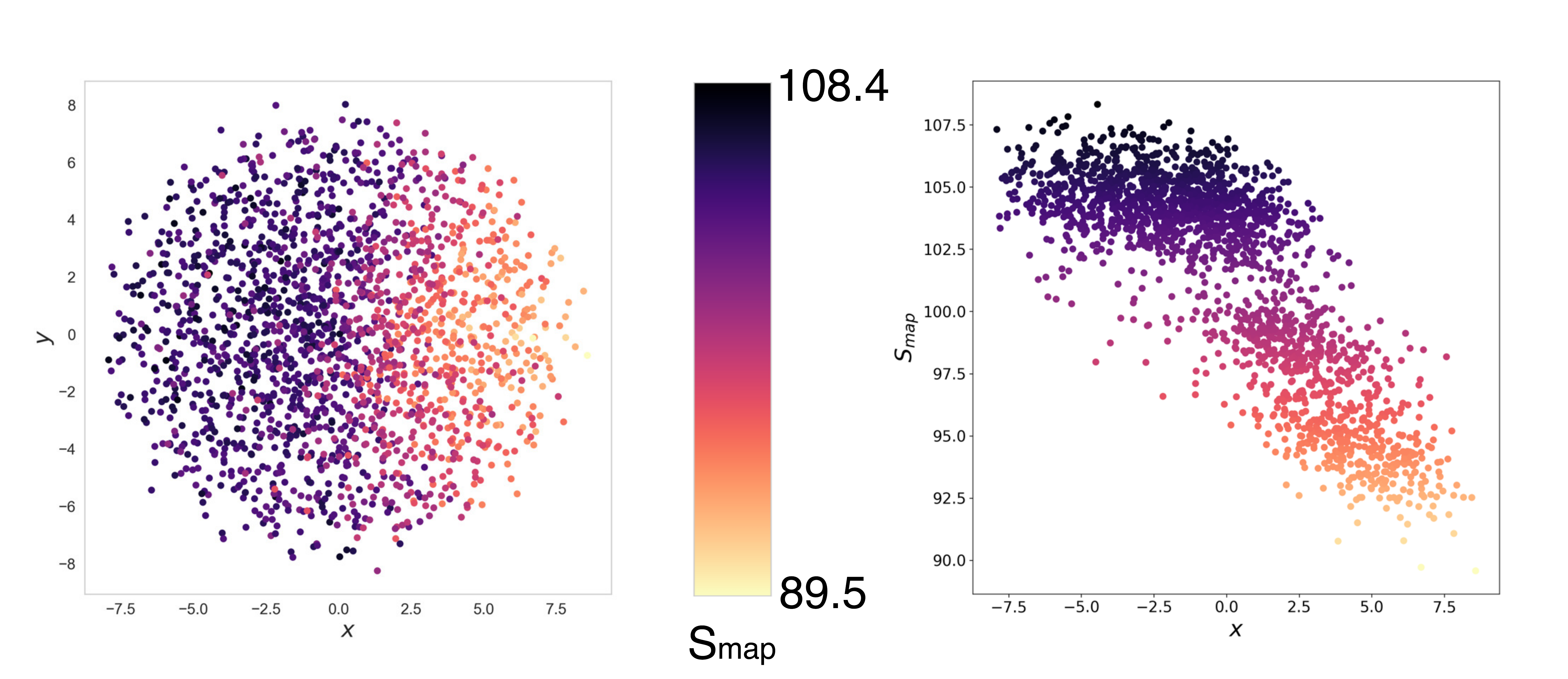}
		\caption{\label{fig:sketch_smap} Application of the sketch map algorithm to a distance matrix obtained calculating $\mathcal{D}$ (Eqs.~\ref{eq:dist_dotmapp} and \ref{eq:rescaled_norm}) over a data set of $1968$ mappings \cite{errica2021deep} that span a wide range of values of mapping entropy. The \textit{x} component separates very well the data points according to their value of $S_{map}$, thus proving that informative mappings can be distinguished among the elements of $\mathcal{M}$ according to a measure of geometrical similarity such as $\mathcal{D}$. The parameters fed to the algorithm are the following:  $\sigma_D = \sigma_d = a_D = b_D = 5$, $a_d = b_d = 2$.}
\end{figure*}

Fig.~\ref{fig:sketch_smap} shows that the two-dimensional embedding obtained through the application of the sketch map algorithm separates the CG mappings according to a gradient of $S_{map}$. In particular, the \textit{x} component of the sketch map and the mapping entropy $S_{map}$ display a clear anticorrelation. The results suggest that highly informative mappings, characterised by low values of $S_{map}$, share geometrical features that are not present in less informative (high $S_{map}$) representations. In other words, the peculiar resolution distribution found in low-$S_{map}$ mappings separates them from the other elements of $\mathcal{M}$. The relevant features that the mapping entropy highlights thus reverberate in the merely structural characterisation provided by the mapping distance; this connection opens the way to an effective usage of the norm $\mathcal E$ and the distance $\mathcal D$ to single out highly informative parts of a macromolecule.

\section{Conclusions}
\label{sec: conclusions}

In this work we have addressed the problem of defining, in a mathematically rigorous manner, the distance between two low-resolution representations of a macromolecule, and to ``explore'' the metric space induced by it.

The recent advances in the computational investigation of soft and biological matter have provided us with the tools to perform large-scale simulations of large and complex systems; however, due to the sheer size of the data produced, one has to filter out the large amount of detail with which the system is described \cite{giulini2021system}, thus relying on a coarse-grained description of it.

Decimation mappings offer a simple and intuitive way of applying this filter, in that only a subset of a molecule's atoms is retained; however, not all mappings entail or deliver the same amount of information, and the identification of the most informative ones allows one to highlight relevant properties of the system. To better understand this relationship between structural representation and physical properties, it is of fundamental importance to possess an instrument to measure the difference among mappings. The metrics proposed here, which builds on the SOAP measure proposed by Cs\'any and coworkers \cite{bartok2013representing,de2016comparing}, has been employed to quantify the number, dissimilarity, and structural features of different mappings, thereby providing the basis for quantitative analysis of the aforementioned relationship.

The exploration of the mapping space relied on the application of the Wang-Landau enhanced sampling algorithm \cite{wang2001determining, wang2001efficient}, which allowed us to compute the (logarithm of the) density of states for mappings with a given number of CG sites, as a function of their squared norm. On the one hand, these calculations brought to the surface information about ``special'' (i.e. atypical) representations that, just due to their lower number with respect to randomly sampled ones, are exponentially suppressed; on the other hand, we made use of the densities of states to implement a lattice-gas analogy in terms of which we have interpreted mappings of qualitatively different types as different phases of the same physical system undergoing a phase transition. Finally, we have made use of the distance between mappings to investigate the properties of optimal reduced representations obtained minimising the mapping entropy, a measure of the amount of information that a given mapping can return about the underlying system at thermal equilibrium: this last analysis has shown that optimal mappings are markedly distant, and therefore qualitatively different, from randomly sampled ones, thus corroborating the idea that the former belong to a particular subregion of the mapping space endowed with nontrivial properties.

A number of questions remain open, that could not be addressed in this work. First, we defined our notion of distance on a single, static structure: it is reasonable to ask if and how to incorporate in it information about the system at thermal equilibrium (or in out-of-equilibrium conditions). Second, the phase transitions that we observed are analogous to the ones observed in a previous work \cite{foley2020exploring}, a connection that certainly deserves to be further inspected. Third, in order to make the most of a simplified description of a system, it is clearly necessary to account for its interactions explicitly, hence a natural next step will be to incorporate them in the exploration of the mapping space.

In conclusion, the mathematical, biophysical, and computational methods developed and applied in this work have served to start gathering the treasure of information buried in the relationship between how we look at a system and the properties it is endowed with, of which we think that what has been reported here has just scratched the surface.

\section*{Data availability}
The raw data produced and analysed in this work are freely available on the Zenodo repository \url{https://doi.org/10.5281/zenodo.4954580}.

\begin{acknowledgments}
The authors thank Giovanni Mattiotti for a critical reading of the manuscript and useful comments. This project has received funding from the European Research Council (ERC) under the European Union's Horizon 2020 research and innovation programme (grant agreement No 758588).
\end{acknowledgments}

\section*{Author contributions}
RP and RM elaborated the study. RM and RP developed the scalar product between mappings. RM developed the software for the mapping norm calculations and the Wang-Landau sampling. MG produced the sketch maps. RM and MG performed the analysis. All authors contributed to the interpretation of the results and the writing of the manuscript.

\appendix

\section{Wang-Landau sampling}
\label{app:wangland}

For each degree of coarse-graining $N$ investigated in this work, see Table~\ref{tab:RSvsSP}, the corresponding density of states $\Omega_N(\mathcal{E})$ defined in Eq.~\ref{eq:omegaE}---that is, the number of possible CG representations in the mapping space $\mathcal{M}$ that retain $N$ atoms and have a squared norm of $\mathcal{E}$---was determined by relying on the protocol proposed by Wang and Landau (WL) \cite{wang2001determining,wang2001efficient,shell2002generalization,barash2017control}.

WL sampling enables to self-consistently determine $\Omega_N(\mathcal{E})$, or, for computational convenience, the associated entropy ${S}_N(\mathcal{E})=\ln[\Omega_N(\mathcal{E})]$, through a sequence $k=0,...,K$ of nonequilibrium Monte Carlo (MC) simulations that provide an increasingly accurate approximation to the correct result \cite{wang2001determining,wang2001efficient}. Given a partition of the ensemble of possible norms $\mathcal{E}$ in bins of width $\delta \mathcal{E}$, the pivotal ingredients of the WL iterative scheme are, respectively: (\emph{i}) the MC estimate of the entropy $\bar{S}_N(\mathcal{E})$; (\emph{ii}) the histogram of visited norms at iteration $k$, $H^k_N(\mathcal{E})$; and (\emph{iii}) the modification factor $\ln(f_k)$ governing convergence of the algorithm---for $k=0$, one typically sets $\bar{S}_N(\mathcal{E})=0$ and $\ln(f_0)=1$.

At the beginning of each iteration $k$, the histogram $H^k_N(\mathcal{E})$ is set to zero. Subsequently, a series of MC moves is performed in which a transition between two mappings $M$ and $M'$, respectively with norms $\mathcal{E}$ and $\mathcal{E}'$, is accepted with probability, see Eq.~\ref{eq:WL_acceptance},
\begin{equation}
\label{eq:WL_MCaccept}
\alpha_{M\rightarrow M'}=\text{min}\left[1,\exp{\left(-[\bar{S}_N(\mathcal{E}')-\bar{S}_N(\mathcal{E})]\right)}\right].
\end{equation}
In our case, both mappings have $N$ sites but differ by the retainment of a single atom. If the move $M\rightarrow M'$ is accepted, the histogram $H^k_N$ and entropy $\bar{S}_N$ are updated according to
\begin{eqnarray}
\label{eq:wl_histoup}
H^k_N(\mathcal{E}')&=&H^k_N(\mathcal{E}')+1,\\
\label{eq:wl_dosup}
\bar{S}_N(\mathcal{E}')&=&\bar{S}_N(\mathcal{E}')+\ln (f_k),
\end{eqnarray}
while in case of rejection one has to replace $\mathcal{E}'$ with $\mathcal{E}$ in Eqs.~\ref{eq:wl_histoup} and~\ref{eq:wl_dosup}. As highlighted by Eqs.~\ref{eq:WL_MCaccept} and~\ref{eq:wl_dosup}, the early stages of the WL scheme tend to ``push away'' the sampling from already visited regions of the mapping space, thus significantly boosting its exploration compared to randomly drawing CG representations. The algorithm then evolves to generate a ``random walk'' in the space of possible norms \cite{barash2017control}.

The series of MC moves within iteration $k$ is interrupted when the histogram of sampled norms $H^k_N(\mathcal{E})$ is ``flat'', meaning that each of its entries does not exceed a threshold distance from the average of the histogram $\langle H^k_N\rangle$. A typical requirement is $p_{flat}\times\langle H^k_N \rangle < H^k_N(\mathcal{E})< (2-p_{flat})\times\langle H^k_N \rangle$ for every value of $\mathcal{E}$, $p_{flat}$ being a predefined flatness parameter. When the flatness condition is satisfied, iteration $k+1$ of the algorithm begins with a reduced modification factor---in our case, we set $\ln (f_{k+1})=\frac{1}{2}\ln(f_{k})$. Finally, iterations over $k$ are stopped when $\ln (f_{k})<\ln (f_{end})\ll 1$, $\ln(f_{end})$ being another control parameter provided in input to the WL protocol. Up to an additive constant, the MC estimate of the entropy $\bar{S}_N(\mathcal{E})$ reproduces the exact result $S_N(\mathcal{E})$ with an accuracy of order $\ln(f_{end})$ \cite{landau2004new}.

In WL sampling, knowledge of the boundaries of the domain of the density of states $\Omega_N(\mathcal{E})$ (equivalently, the entropy $S_N$) plays a crucial role in the convergence the iterative scheme, e.g., for checking the flatness of the histogram $H_N(\mathcal{E})$ throughout the simulation \cite{wust2008hp}. In contrast to ``more traditional'' systems such as Ising ferromagnets on a lattice \cite{wang2001determining}, this information is not readily available in our case. As such, we initially performed a set of explorative, non-iterative---i.e., without updating the modification factor $\ln(f_{k})$---WL runs so as to approximately locate the minimum and maximum norms $\mathcal{E}_{min}(N)$ and $\mathcal{E}_{max}(N)$ achievable at each degree of CG'ing. To mitigate the effect of bins that are only visited at a very late stage of the simulation, thus risking to temporarily ``trap'' the mapping space exploration, we followed the protocol described in Ref.~\cite{wust2008hp}: every time a bin $[\mathcal{E}_i,\mathcal{E}_i+\delta\mathcal{E}]$ was populated for the first time, it was marked as ``visited'', the corresponding entropy was initialised to the minimum of $\bar{S}_N({\mathcal{E}})$ over the previously visited bins, and the histogram $H_N(\mathcal{E})$ was reset. The results obtained from these preliminary runs for $\mathcal{E}_{min}(N)$ and $\mathcal{E}_{max}(N)$ as a function of $N$ are displayed in Fig.~\ref{fig:scaling_ran_WL} and summarised in Table~\ref{tab:WLwindef}.

\begin{figure*}
		\includegraphics[width=\textwidth]{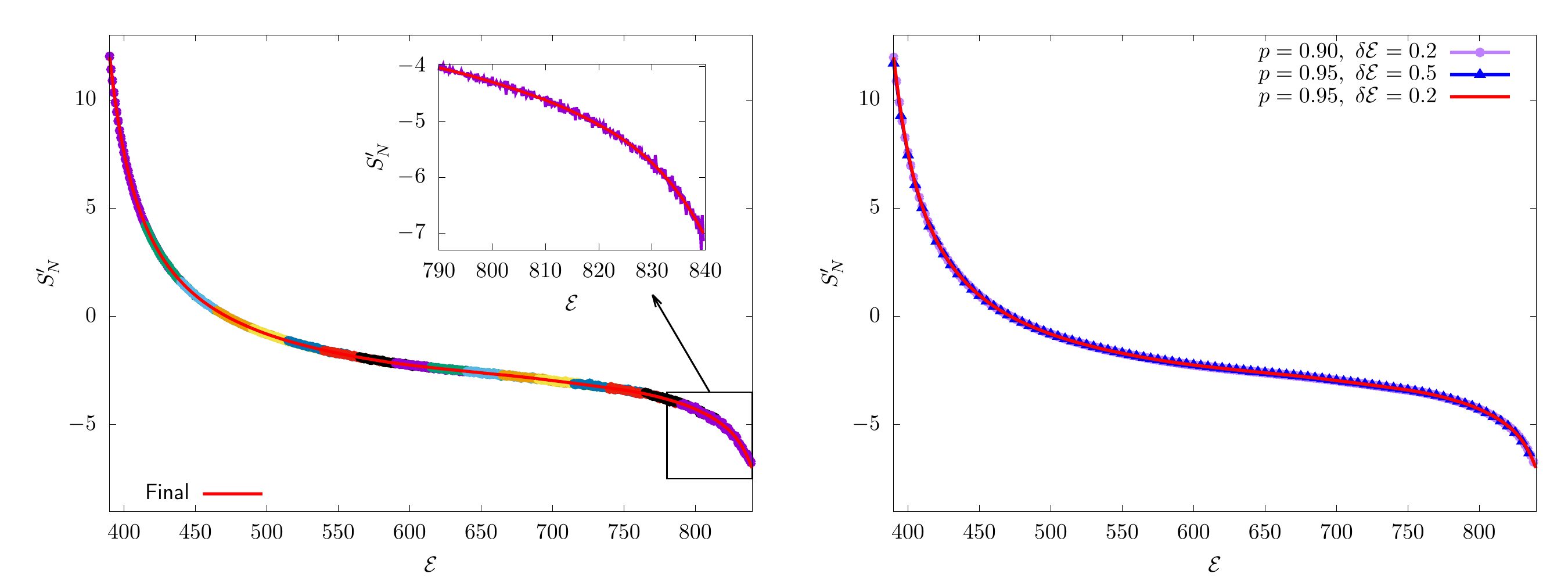}
		\caption{\emph{Left}: Main figure: Comparison between the piecewise entropy derivatives $S'_{N,i}(\mathcal{E})$, $i=1,...,W_N$ of AKE (colored dots) obtained from the set of independent WL simulations performed over the $W_N$ overlapping windows, and the final, reconstructed derivative $S'_N$ (``Final'', red line) calculated through the mixing procedure of the $S'_{N,i}$ described in the text. We report results for $N=856$. Inset: Behaviour of the derivative $S'_{N,i}(\mathcal{E})$ for the last WL window before and after the application of the Saviztky-Golay filter to the raw simulation results for the entropy $\bar{S}_{N,i}(\mathcal{E})$. \emph{Right}: Reconstructed derivates $S'_{N}(\mathcal{E})$ for $N=856$ obtained by varying a subset of the input parameters of the WL protocol. Specifically, we test the sensitivity of the results to a change in the flatness parameter $p_{flat}$ as well as in the bin width $\delta\mathcal{E}$, considering as reference profile the derivative $S'_N$ obtained by setting $p_{flat}=0.95$ and $\delta\mathcal{E}=0.2 $ (red full line).\label{fig:derivative_smooth}}
\end{figure*}

\begingroup
\setlength{\tabcolsep}{9pt}
\renewcommand{\arraystretch}{1.1}
\begin{table}[tp]
\begin{tabular}{c c c c c c}
\hline
 N & $\mathcal{E}_{min}$ & $\mathcal{E}_{max}$ & $\bar{\mathcal{E}}_{min}$ & $\bar{\mathcal{E}}_{max}$ & $W_N$\\
\hhline{======}
214 & 25.6 & 209.8 & 28 &192 & 3\\
428 & 92.4 & 424.4 & 98 &410 & 7\\
642 & 206.2 & 633.8 & 218 & 618 & 15\\
856 & 371.2 & 853.0 & 390 &840 & 17\\
1070 & 600.2 &1074.2 & 612 & 1062 & 17\\
1284 & 900.2 & 1298.4 & 910 &1290 & 18\\
1498 & 1287.8 & 1514.6 & 1296 &1504 & 12\\
\hline
\end{tabular}
\caption{\label{tab:WLwindef} Lower and upper bound of the norms $\mathcal{E}_{min}$ and $\mathcal{E}_{max}$ identified by the set of preliminary WL runs for each degree of CG'ing $N$, see Fig.~\ref{fig:scaling_ran_WL}, and corresponding values $\bar{\mathcal{E}}_{min}$ and $\bar{\mathcal{E}}_{max}$ employed in the reconstruction of the entropy $S_N(\mathcal{E})$ through the iterative WL scheme. For boosting convergence of the algorithm, the interval $[\bar{\mathcal{E}}_{min},\bar{\mathcal{E}}_{max}]$ was divided in $W_N$ windows overlapping by half their width; the associated simulations were performed with a flatness parameter $p_{flat}=0.90$, assuming convergence of the iterations when the modification factor $\ln(f_{k})$ became smaller than $\ln(f_{end})=10^{-6}$.}\end{table}
\endgroup

Having identified the range of possible norms for each investigated degree of CG'ing, we subsequently moved to the determination of the corresponding entropies $S_N(\mathcal{E})$ via the iterative WL scheme. To boost convergence of the algorithm, for each $N$ we slightly reduced the interval of norms $[\mathcal{E}_{min}(N),\mathcal{E}_{max}(N)]$ with respect to the one predicted by the explorative runs, and divided this spectrum in a total of $W_N$ overlapping windows of equal width, see Table~\ref{tab:WLwindef} \cite{wang2001efficient}. The overlap between two consecutive windows was fixed to half their size. Within each window, we then performed a separate WL simulation in which confinement of the range of norms was achieved by rejecting all mapping moves $M\rightarrow M'$ that would bring the exploration outside the $\mathcal{E}$ interval of interest. In discarding these moves, we concurrently updated the histogram and entropy of the current state according to Eqs.~\ref{eq:wl_histoup} and~\ref{eq:wl_dosup} in order to avoid boundary effects \cite{schulz2003avoiding}. Furthermore, also in these production runs we kept track of the norm bins that were sampled during the course of the simulation, resetting the histogram every time a new bin was populated, the entropy of which was initialised to the minimum of $\bar{S}_N({\mathcal{E}})$ over the previously visited ones. All WL simulations were performed setting $p_{flat}=0.90$, and checking the histogram flatness over the visited bins every $3\cdot 10^6$ ``single spin'' MC moves that involved the swap of a retained and a non-retained atom in the mapping. We interrupted the iterative scheme  when the modification factor $\ln(f_{k})$ became smaller than $\ln(f_{end})=10^{-6}$.

For each degree of CG'ing $N$, the outcome of the converged WL protocol is a set of entropies $\bar{S}_{N,i}(\mathcal{E})$, $i=1,...,W_N$, restricted to bounded and overlapping $\mathcal{E}$ domains that need to be combined to provide the complete $S_N(\mathcal{E})$ over the whole range of investigated norms. These $\bar{S}_{N,i}(\mathcal{E})$ differ---besides numerical uncertainties that are inherent to the self-consistent scheme \cite{belardinelli2007fast}---from the exact results $S_{N,i}(\mathcal{E})$ by additive constants $C_{N,i}$ that are not uniform across the different WL windows. Rather then determining the relative shifts that most accurately superimpose the various $\bar{S}_{N,i}(\mathcal{E})$ profiles within the overlapping regions---see e.g. Ref.~\cite{shell2002generalization}---in this work we directly considered the (numerical) derivatives of $\bar{S}'_{N,i}(\mathcal{E})$ in each WL window,
\begin{equation}
\bar{S}'_{N,i}(\mathcal{E})=\frac{d \bar{S}_{N,i}(\mathcal{E})}{d\mathcal{E}}=\frac{1}{T},
\end{equation}
where $T$ is the ``temperature'' of the system. These derivatives are not affected by the constants $C_{N,i}$, so that each $\bar{S}'_{N,i}(\mathcal{E})$ is approximately equal to its exact counterpart $S'_{N,i}(\mathcal{E})$. One can thus combine all the derivatives of the different WL windows in a global derivative $S'_N(\mathcal{E})$ that extends over the whole range of analysed norms, from which the overall entropy $S_N(\mathcal{E})$ can be calculated as
\begin{equation}
\label{eq:entropy_integ}
S_N(\mathcal{E})=S_N(\mathcal{E}_{min}(N))+\int_{\mathcal{E}_{min}(N)}^{\mathcal{E}}S'_N(\mathcal{E'})d\mathcal{E'},
\end{equation}
where $\mathcal{E}_{min}(N)$ is the lowest norm sampled at degree of CG'ing $N$. Note that in contrast to systems as the ferromagnetic Ising model, we do not \emph{a priori} know the value of $S_N(\mathcal{E}_{min})$, so that the entropy $S_N(\mathcal{E})$ will be only determined up to a constant. 

To merge the set of derivatives and reconstruct $S'_N(\mathcal{E})$ for each degree of CG'ing, we first applied a Savitzky-Golay filter \cite{savitzky1964smoothing} to the WL estimates of the entropies $\bar{S}_{N,i}(\mathcal{E})$ so as to reduce the amount of noise in the simulation results, and consequently smoothen the derivative $S'_{N,i}(\mathcal{E})$ of each window. A comparison of the derivatives obtained in presence or absence of the filter, see Fig.~\ref{fig:derivative_smooth}, highlights how this only applies a tiny correction to the original data, which nonetheless significantly improves the quality of the set of $S'_{N,i}(\mathcal{E})$. Despite this refinement, the presence of residual numerical fluctuations leave room to a certain degree of arbitrariness in how, within the overlap region of two consecutive windows, the combined derivative should be constructed. At the same time, these fluctuations appear to be marginal in the vicinity of the center of a window, while tend to slightly increase if we move towards its boundaries (data not shown). Exploiting this observation, we thus tackled the problem of merging the derivatives of two consecutive windows $i$ and $i+1$ within their overlap region as follows: first, we divided the region in three separate intervals, the central one being roughly double the size of the other two. Given that the windows overlap by half their width, it follows that the first interval will be located close to the center of window $i$, where the derivative $S'_{N,i}(\mathcal{E})$ is numerically more stable, but close to the boundary of window $i+1$, where $S'_{N,i+1}(\mathcal{E})$ is slightly more noisy. The opposite holds for the last interval. As such, in the first and last region we considered the combined derivative $S'_N(\mathcal{E})$ to be equal to $S'_{N,i}(\mathcal{E})$ and $S'_{N,i+1}(\mathcal{E})$, respectively. 
Within the central interval, by increasing $\mathcal{E}$ we move from the vicinity of the center of window $i$ to that of window $i+1$. In this latter region we thus set the final derivative $S'_N$ to a weighted average of the derivatives of the two windows, namely
\begin{equation}
S'_N(\mathcal{E})=(1-\alpha(\mathcal{E}))S'_{N,i}(\bar{\mathcal{E}})+\alpha(\mathcal{E})S'_{N,i+1}(\mathcal{E}),
\end{equation} 
where $\alpha(\mathcal{E})$ is a mixing parameter that linearly increases from zero to one as $\mathcal{E}$ moves from the left to the right boundary of the interval. 

\begin{figure*}
		\includegraphics[width=\textwidth]{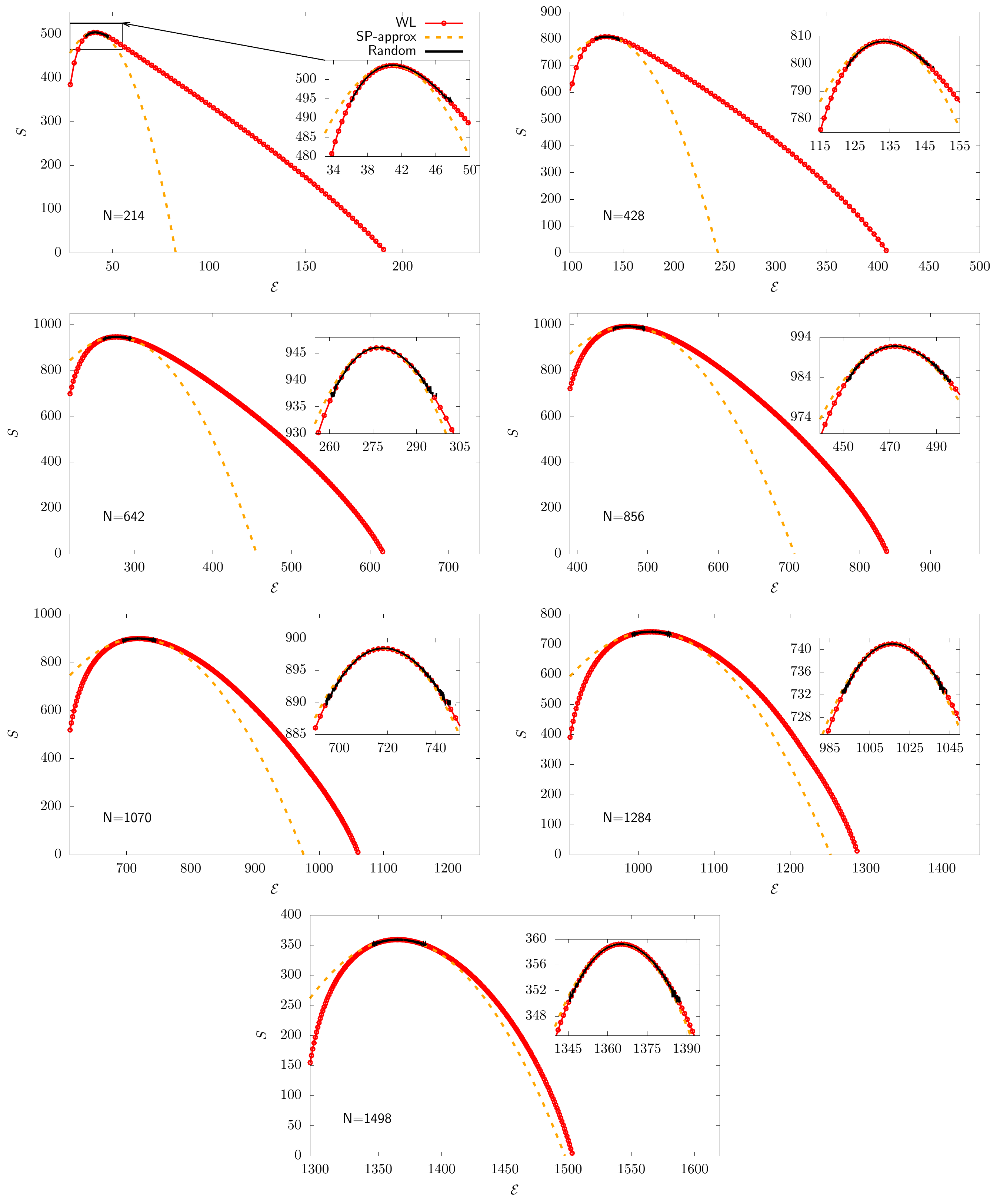}
		\caption{Behavior of the entropy $S_{N}(\mathcal{E})$ of AKE for different degrees of CG'ing. For each $N$, we report results obtained via (\emph{i}) Wang-Landau sampling (``WL'', red dotted lines), shifting the data so that the minimum of $S_N$ over the range of investigated norms is zero;  (\emph{ii}) a saddle-point approximation of the WL predictions (``SP-approx'', orange dashed lines); and (\emph{iii}) a random drawing of CG representations (``Random'', black lines), in this latter case shifting the curve so that its maximum coincides with the one of the corresponding WL profile. \label{fig:dos_N_WL}}
\end{figure*}

Repeating this interpolation for all the set of $W_N$ windows---note that in the first (resp. last) half of the first (resp. last) window no mixing applies---provided us, for each of the analysed degrees of CG'ing, with a global derivative $S'_N(\mathcal{E})$ that extends over the whole range of sampled norms. Fig.~\ref{fig:derivative_smooth} displays a comparison between $S'_N(\mathcal{E})$ and the original, piecewise derivatives for the case $N=856$, highlighting the accuracy of our approach. This accuracy is further confirmed by the smooth behavior of the second derivative $S''_N(\mathcal{E})$ calculated from the reconstructed $S'_N$, that we display in Fig.~\ref{fig:entropy_856} for $N=856$. Starting from the set of $S'_N(\mathcal{E})$, the corresponding entropies $S_N(\mathcal{E})$ were subsequently obtained via direct integration, see Eq.~\ref{eq:entropy_integ}, producing the profiles presented in Fig.~\ref{fig:entropy_856} and in Fig.~\ref{fig:dos_N_WL}. In these figures, entropies were shifted so that their minimum value is zero.

Finally, it is interesting to test the dependence of our results on the input parameters of the WL protocol. While initially all MC simulations were performed with a flatness condition $p_{flat}=0.90$, for the case $N=856$ we repeated the calculations using $p_{flat}=0.95$ finding a perfect agreement of the reconstructed $S'_N({\mathcal E})$, see Fig.~\ref{fig:derivative_smooth}. The same sensitivity analysis was performed for the bin size $\delta\mathcal{E}$ dictating the discretisation of the mapping norms: while in all simulations we employed $\delta\mathcal{E}=0.2$, by repeating the calculations for $N=856$ with a bin width of $\delta\mathcal{E}=0.5$ we again observed excellent agreement of the results, see Fig.~\ref{fig:derivative_smooth}.

\section{Heat capacity of the lattice gas}
\label{app:latticegas}

\begin{figure*}
		\includegraphics[width=\textwidth]{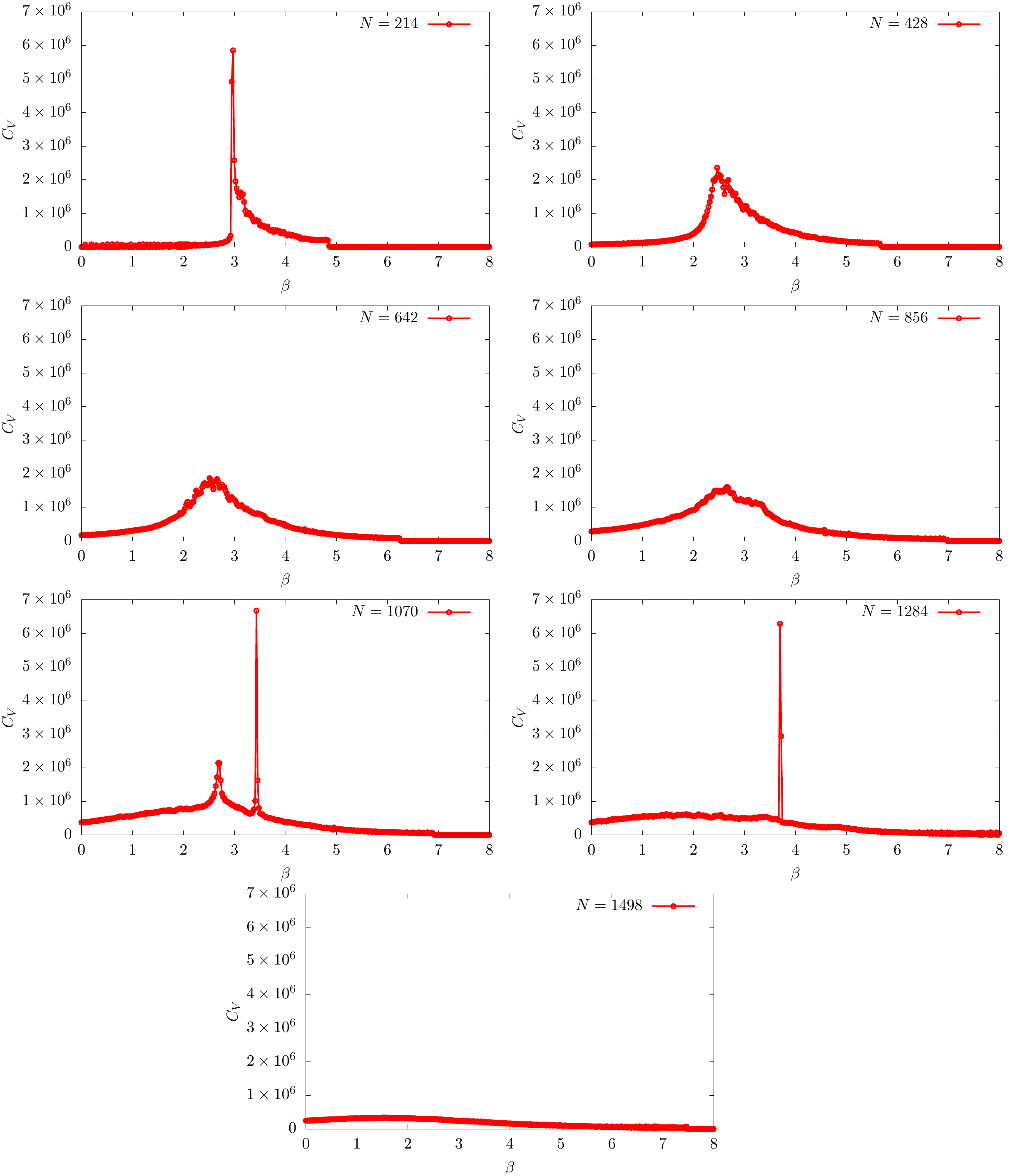}
		\caption{Dependence of the heat capacity $C_V$ on the inverse temperature $\beta$ for the lattice gas analogue of the mapping norm of AKE calculated at several degrees of CG'ing. Sharp peaks in $C_V$ at high (resp. low) values of $N$ suggest the presence of a solid-liquid (resp. liquid-gas) transition in the system. It should be noted that the scales of $\beta$ and $C_V$ are the same in all plots. \label{fig:spec_heat}}
\end{figure*}

In this Appendix we provide additional information about the phase transitions observed in the lattice gas analogue of the mapping norm. Specifically, Fig.~\ref{fig:spec_heat} displays the heat capacity $C_V$ of the lattice gas, see Eq.~\ref{pt:7} in the main text, as a function of the inverse temperature $\beta$ for different degrees of CG'ing $N$, calculated from the Legendre-Fenchel transform $\beta F_N(\beta)$ of the WL entropies $S_N(E)$, see Eq.~\ref{pt:4}. While for the highest degree of CG'ing investigated, $N=1438$, the heat capacity has a smooth dependence on $\beta$, for $N=1070$ $C_V$ develops a sharp peak for low temperatures, which suggests the presence of a solid-liquid transition in the system. By further decreasing the degree of CG'ing, this solid-liquid peak gets initially flanked by a shoulder located at higher temperatures, and finally disappears. The shoulder, on the other hand, grows in magnitude as fewer and fewer sites are retained, and becomes a discontinuity for $N=214$, suggesting the appearance of a liquid-gas transition.\newline

\begin{figure*}
		\includegraphics[width=\textwidth]{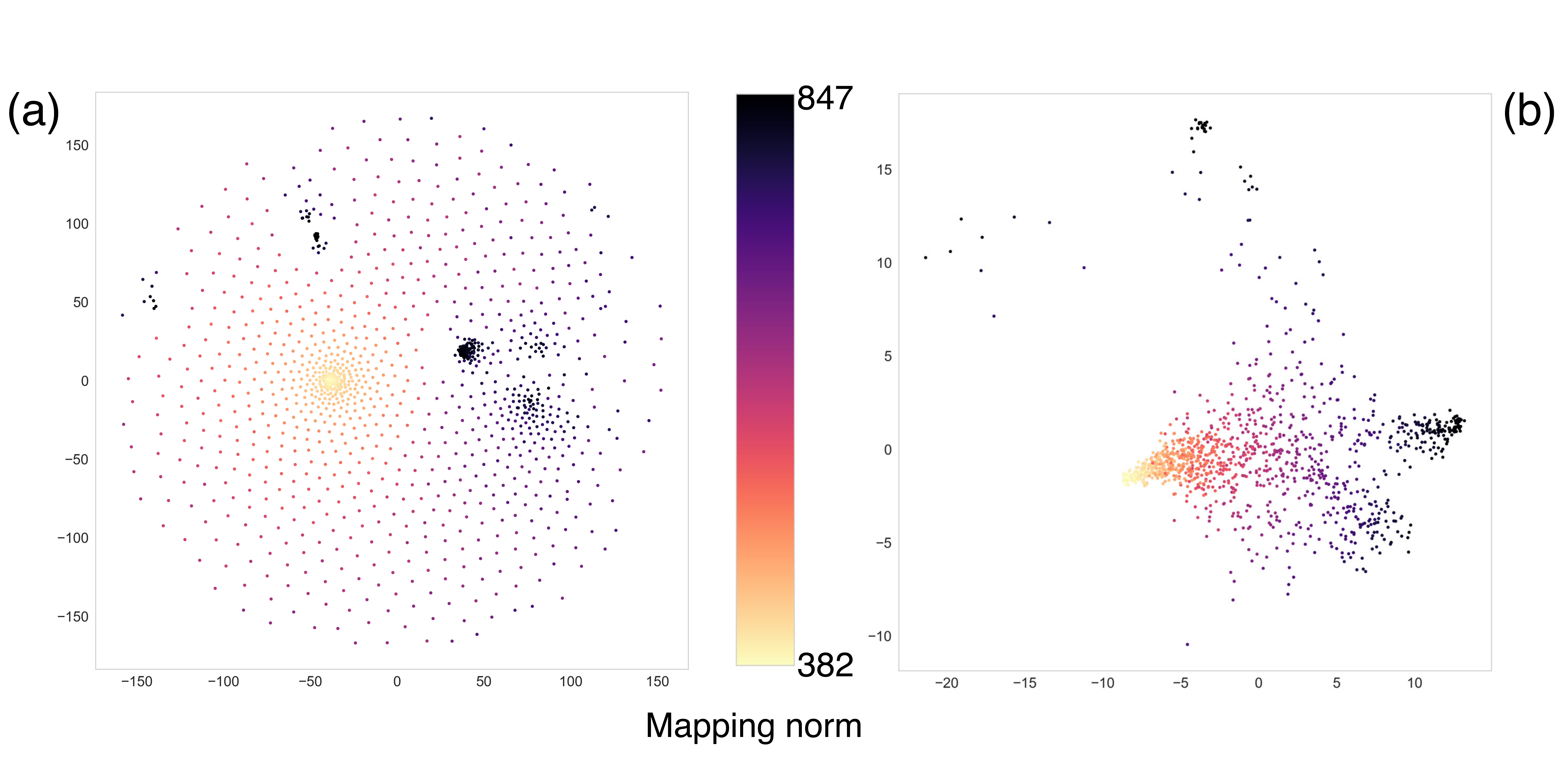}
		\caption{\label{fig:sketch_map_sigma25} Application of the sketch map algorithm to the mapping space $\mathcal{M}$: case of $N = 856$. We employed the same set of parameters described in Fig. \ref{fig:sketch} of the main text, where CG mappings have $N=214$, with the exception of $\sigma_D$ and $\sigma_d$ in subfigure (a), which are equal to $5$. The two-dimensional embedding shown here displays similar properties to the one in the main manuscript; specifically, if $\sigma_D$ and $\sigma_d$ have low values, essentially all the data points are depicted as isolated instances in  $\mathcal{M}$ and only the extremely sparse and globular mappings are capable of forming recognisable clusters. With a higher value of these parameters, all sparse mappings collapse in a well-defined region of the plane, from which several \textit{routes} depart, each one directed towards globular mappings covering different domains of the protein structure.}
\end{figure*}

\bibliographystyle{ieeetr}
\bibliography{main.bib}

\end{document}